\documentclass[sigconf]{acmart}
\usepackage{hyperref}
\usepackage{longtable}
\usepackage{seqsplit}
\usepackage{enumitem}
\usepackage{xcolor}
\usepackage{placeins}
\usepackage{makecell}
\PassOptionsToPackage{final}{graphicx}

\definecolor{ElectricPurple}{RGB}{191, 0, 255}
\definecolor{MintGreen}{RGB}{62, 180, 137}
\definecolor{CoralPink}{RGB}{255, 127, 127}
\definecolor{DeepSkyBlue}{RGB}{0, 191, 255}
\definecolor{SunsetOrange}{RGB}{255, 94, 77}
\definecolor{CyberYellow}{RGB}{255, 211, 0}
\definecolor{AquaCyan}{RGB}{0, 255, 200}
\definecolor{RoseGold}{RGB}{183, 110, 121}
\definecolor{MidnightBlue}{RGB}{25, 25, 112}
\definecolor{NeonGreen}{RGB}{57, 255, 20}
\definecolor{Slate}{RGB}{112, 128, 144}

\newcommand{\orchestration}[1]{\textls[0]{\textbf{\textsc{\textcolor{MidnightBlue}{Orchestration}}}}}
\newcommand{\creation}[1]{\textls[0]{\textbf{\textsc{\textcolor{MintGreen}{Creation}}}}}
\newcommand{\insight}[1]{\textls[0]{\textbf{\textsc{\textcolor{SunsetOrange}{Insight}}}}}

\usepackage{booktabs,multirow,tabularx,array}
\usepackage[table]{xcolor}

\AtBeginDocument{%
  }

\copyrightyear{2026}
\acmYear{2026}
\setcopyright{cc}
\setcctype{by}
\acmConference[CAIS '26]{ACM Conference on AI and Agentic Systems}{May 26--29, 2026}{San Jose, CA, USA}
\acmBooktitle{ACM Conference on AI and Agentic Systems (CAIS '26), May 26--29, 2026, San Jose, CA, USA}
\acmDOI{10.1145/3786335.3813140}
\acmISBN{979-8-4007-2415-2/2026/05}

\begin{document}

\title{Why Johnny Can’t Use Agents: Industry Aspirations vs. User Realities with AI Agents}

\author{Pradyumna Shome}
\email{pradyumnashome@cmu.edu}
\orcid{0000-0003-2768-9103}
\affiliation{
  \institution{Carnegie Mellon University}
  \city{Pittsburgh}
  \state{Pennsylvania}
  \country{USA}
}
\author{Sashreek Krishnan}
\email{sashreek@andrew.cmu.edu}
\orcid{0009-0003-0655-2754}
\affiliation{
  \institution{Carnegie Mellon University}
  \city{Pittsburgh}
  \state{Pennsylvania}
  \country{USA}
}
\author{Sauvik Das}
\email{sauvik@cmu.edu}
\orcid{0000-0002-9073-8054}
\affiliation{
  \institution{Carnegie Mellon University}
  \city{Pittsburgh}
  \state{Pennsylvania}
  \country{USA}
}

\renewcommand{\shortauthors}{Shome et al.}

\begin{abstract}
There is growing imprecision about what ``AI agents'' are, what they can do, and how effectively they can be used by their intended users. We pose two key research questions: (i) How does the tech industry conceive and market ``AI agents''? (ii) What challenges do end-users face when attempting to use commercial AI agents for their advertised uses?  We first performed a systematic review of marketed use cases for 102 commercial AI agents, finding that they fall into three umbrella categories: orchestration, creation, and insight. We then evaluated whether end-users could realize these marketed capabilities in practice: we conducted a usability assessment where $N = 31$ participants attempted representative tasks for each of these categories on two popular commercial AI agent tools: Operator and Manus. We found that users were generally impressed with these agents but faced significant usability challenges ranging from agent capabilities that were misaligned with user mental models to agents lacking the meta-cognitive abilities necessary for effective collaboration.
\end{abstract}

\begin{CCSXML}
<ccs2012>
   <concept>
       <concept_id>10003120.10003123.10011759</concept_id>
       <concept_desc>Human-centered computing~Empirical studies in interaction design</concept_desc>
       <concept_significance>500</concept_significance>
       </concept>
   <concept>
       <concept_id>10003120.10003121.10003122.10003334</concept_id>
       <concept_desc>Human-centered computing~User studies</concept_desc>
       <concept_significance>500</concept_significance>
       </concept>
   <concept>
       <concept_id>10003120.10003121.10003122.10010854</concept_id>
       <concept_desc>Human-centered computing~Usability testing</concept_desc>
       <concept_significance>500</concept_significance>
       </concept>
   <concept>
       <concept_id>10003120.10003121.10003124.10010870</concept_id>
       <concept_desc>Human-centered computing~Natural language interfaces</concept_desc>
       <concept_significance>500</concept_significance>
       </concept>
 </ccs2012>
\end{CCSXML}

\ccsdesc[500]{Human-centered computing~Empirical studies in interaction design}
\ccsdesc[500]{Human-centered computing~User studies}
\ccsdesc[500]{Human-centered computing~Usability testing}
\ccsdesc[500]{Human-centered computing~Natural language interfaces}

\keywords{AI Agents, Usability Studies, Interface Agents, LLMs}

% \begin{teaserfigure}
%       \includegraphics[width=1\linewidth]{figures/Teaser.pdf}
%       \caption{We identify five critical usability barriers novice users face when interacting with AI agents to accomplish a task. As they attempt to delegate their task, they must progressively develop a mental model of the agent's capabilities, navigate its rigid collaboration style, decode its overwhelming communication, all while the agent isn't fully cognizant of its own capabilities, and presumes it has a user's complete trust.}
%       \label{fig:teaser}
% \end{teaserfigure}

\maketitle

\section{Introduction}
% Visions of computing systems as intelligent partners for knowledge work stretch back nearly a century with Vannevar Bush's 1945 dream of the ``Memex,'' a mechanized desk that could augment human memory and association by linking together trails of knowledge \cite{memex_as_we_may_think}.
The AI and HCI communities have long been animated by visions of computing systems as intelligent partners for knowledge work: from Vannevar Bush's 1945 dream of the ``Memex'' to Horvitz's explorations of agent-mediated ``mixed-initiative'' interfaces~\cite{principles_of_mixed_initiative_user_interfaces} to Shneiderman and Maes' canonical debate on direct manipulation versus interface agents~\cite{maes1994, shneiderman_maes_1997}.
% Along the way, ill-fated counter examples such as Microsoft's Clippy~\cite{clippy} highlighted the risks of overzealous agent designs that, to paraphrase Adar, had UIs that wrote checks that the underlying AI couldn't cash~\cite{adar2018bounced}.
Today, the dream has resurfaced under the banner of large language model (LLM) powered ``AI agents''.
But are these agents, at long last, useful and usable?
% --- large language model (LLM)-powered systems that autonomously execute multi-step tasks on behalf of users, taking actions with real-world consequences such as booking flights, sending emails, and publishing content. 
% Unlike chatbots that use LLMs to generate natural language responses to user prompts, ``agents'' directly act on behalf of users, promising to orchestrate workflows, generate creative outputs, and deliver insights on demand.
% These properties are what we mean by ``agentic'' throughout; our findings target interaction failures that arise from them, not from LLM accuracy or single-turn dialogue quality alone.

Answering that question is complicated because what an ``AI agent'' is, what it can do, and how those definitions shape user expectations is imprecise and rapidly evolving. Unlike chatbots that use LLMs to generate natural language responses to user prompts, prior work suggests that \emph{operationally agentic} AI agents should have at least some of the following properties: (i) \emph{delegation and initiative allocation}---deciding who between the human user and the AI decides what, when~\cite{principles_of_mixed_initiative_user_interfaces}; (ii) \emph{acting in an environment} rather than only producing text---invoking tools, APIs, or GUIs on the user's behalf~\cite{Lieberman1997Autonomous, shneiderman_maes_1997, maes1994}; (iii) \emph{multi-step execution} with a reason-act loop (e.g., interleaved reasoning and tool use~\cite{yao2023react, schick2023toolformer}); and (iv) \emph{oversight, intervention, and recovery}---the need for users to supervise and correct ongoing agent activity~\cite{sheridan1992telerobotics, mozannar2025magenticui}. 
But, as with other technologies at the peak of their hype cycle, there are also many \emph{marketed-as-agentic} products that self-label as ``AI agents'' that may not fully instantiate any of these properties.
Because general end-users may not understand the difference, in this paper we consider both scopes as relevant.

% what an ``AI agent'' is, what it can do, and how well it fits into the messy realities of everyday knowledge work remain under-specified.
% To that end, beyond the \emph{operationally agentic} scope, we also explore two scopes in this paper: \emph{marketed-as-agentic} products are those that self-label or are described as ``agentic''/``agent'' in marketing or aggregators that may or may not instantiate these operational properties; \emph{operationally agentic} systems are those that do instantiate them.

% While thousands of start-ups and major technology companies claim that these operationally- or marketed-as- agentic can be productivity multipliers for general end-users
To date, much of the conversation about agent performance --- in benchmarks, demo videos, venture funding pitches, and media hype --- has focused on easy-to-quantify ideals that overestimate true effectiveness \cite{zhu2025establishing}.
% But does an agent scoring 26\% versus 24\% on ``Humanity’s Last Exam'' matter to a user when a user wants help with booking a flight?
% Relative to standard LLM-based interactions, usability failures for agentic AI products can be more immediately costly: a poorly designed chatbot produces a bad answer, but a poorly designed agent books the wrong flight.
To better design and evaluate ``AI agents'' that are useful and usable for end-users, we need a systematic understanding of how the tech industry conceives of ``AI agents'', how those framings shape user expectations, and where end-users' lived experience diverges from marketed aspirations. 
Against this backdrop, we ask:
% \paragraph{Research Questions}

\begin{enumerate}[label=\textbf{RQ\arabic*:}]
    \item How does the tech industry conceive of and market ``AI agents''? Industry framings matter because they define the menu of software choices that most users can access and set the expectations with which users approach these tools.
    \item What challenges do end-users face when using commercial AI agents for their advertised uses? We seek to understand the frictions that emerge when the aspirational capabilities these products imply collide with the realities of use.
\end{enumerate}

% \paragraph{Summary of Work}

% \begin{figure*}[ht]
%     \centering
%     \includegraphics[width=1\linewidth]{figures/Research-Methodology.pdf}
%     \caption{Overview of our research process. We conducted a systematic review to build a taxonomy of marketed use cases of AI agents. Next, we conducted a think-aloud session in which participants completed tasks corresponding to our umbrella use cases, from which we identified usability barriers. These barriers lead to a set of design implications for next-generation AI agent design.}
%     \label{fig:overview-of-methodology}
% \end{figure*}

To answer RQ1, we constructed a taxonomy of \emph{marketed-as-agentic} products. We started by sourcing advertised uses for AI agents by exploring aggregator websites for AI Agent products, including the AI Agents Directory and Product Hunt, and supplemented our corpus with web searches to increase the diversity of products. We then analyzed themes across verbiage relating to agent functionalities and distilled a taxonomy of three broad advertised use-cases for AI agents: \orchestration{}, \creation{}, and \insight{}. 
% Figure ~\ref{fig:overview-of-methodology} provides an overview of our research process.

To answer RQ2, we ran a think-aloud usability study in which $N=31$ participants attempted representative tasks in each of the three categories using two popular \emph{operationally agentic} platforms---Operator and Manus. We supplemented these think-aloud studies with semi-structured interviews. Although participants were generally successful at accomplishing the assigned tasks (which were specifically selected to be simple and doable), participants struggled with five critical usability barriers:
(i) agent capabilities are misaligned with user mental models, (ii) agents presume trust without establishing credibility, (iii) agents fail to accommodate a diversity of collaboration styles, (iv) agents generate an overwhelming amount of communication overhead, and (v) agents lack the meta-cognitive abilities that enable productive collaboration.
% Figure ~\ref{fig:teaser} symbolically depicts Johnny, our proverbial end-user, navigating these challenges as they attempt to delegate a task. 

% These usability breakdowns echo long-standing critiques of agentic systems in HCI: the dangers of ceding too much initiative, the costs of opaque automation, and the challenge of aligning agent actions with human goals.

% \paragraph{Contributions}

In summary, this paper makes three core contributions.
(1) We introduce a taxonomy of marketed AI agent capabilities, distilling how the tech industry conceives of and frames AI agents.
(2) We present an empirical account of five critical usability barriers end-users face when attempting to use these tools on a representative set of tasks. Relative to prior human-AI interaction work, we contribute (a) the expectations-to-breakdowns chain for agentic systems, (b) a mapping of usability barriers to operational agentic properties, and (c) evaluation dimensions for delegation, oversight, and recovery in agentic systems.
(3) We distill design and evaluation implications for agentic systems that are usable and effective for end-users.
\section{Related Work}
Our work is situated at the intersection of three areas of inquiry in the HCI and AI literature: the design and evolution of interface agents, evaluations of AI agents and LLMs, and empirical studies of human-AI interaction.
% Prior work has envisioned agents through conceptual frameworks and limited early deployments that predate today's autonomous AI agents, studied human-AI collaboration with less autonomous systems such as chatbots and decision-support tools, and evaluated agent capabilities through synthetic benchmarks rather than usability with real users on everyday tasks. Although agent interfaces may superficially resemble those of chatbots, the autonomous actions agents take on behalf of users amplify the consequences of usability failures, demanding a higher standard of interaction design.
% We contribute an empirical evaluation of widely-deployed commercial AI agents with real users, finding that usability barriers --- not capability limitations --- are the primary bottleneck to end-users realizing agent potential.

\textbf{Design and Evolution of Interface Agents.}
\textit{Direct manipulation} has long been the dominant interaction paradigm for GUI interfaces ~\cite{shneiderman1983direct}.
However, \textit{interface agents} that act as intermediaries and invoke commands on users' behalf~\cite{maes1994} have been touted as a complementary, and at times oppositional, paradigm.
This tension --- crystallized in the classic debate over direct manipulation versus interface agents~\cite{maes1994, shneiderman_maes_1997} and the broader tradition of mixed-initiative interfaces~\cite{principles_of_mixed_initiative_user_interfaces} --- has motivated decades of work exploring agent behaviors, interaction styles, and control mechanisms~\cite{LiebermanSelker1999,Lieberman1997Autonomous}. Researchers have also used Wizard-of-Oz methods to prototype agent experiences beyond the limits of contemporary AI~\cite{maulsby1993wizard}, while commercial assistants from Microsoft Clippy to voice assistants have made the promise (and pitfalls) of agent-mediated interaction visible to end-users~\cite{clippy, Alexa, Siri, Cranshaw2017CalendarHelp}.

Recent transformer-based LLMs (e.g., ~\cite{gpt3}) have renewed excitement for interface agents by enabling GUI agents that plan and execute multi-step actions in real software environments~\cite{hu2025osagents, zhang2025largelanguagemodelbrainedgui}.
Yet, many critical usability questions remain: e.g., what autonomy should agents take, how should tasks be delegated, and how should oversight mechanisms be designed to keep users in control~\cite{sheridan1992telerobotics, maes1994}?
We contribute an empirical examination of commercially deployed LLM-powered AI agents with real users, identifying the usability barriers that persist in today’s renewed instantiation of interface agents and that limit effective delegation, oversight, and collaboration.

\textbf{Evaluations of LLMs and AI Agents.}
A broad ecosystem of benchmarks and evaluation frameworks now exists for evaluating AI agents: e.g., capability benchmarks~\cite{chiang2024chatbot, rawles2024androidworld, swe_bench, phan2025humanitysexam}, evaluation environments for computer-based knowledge work~\cite{osworld, drouin2024workarenacapablewebagents, pan2024webcanvasbenchmarkingwebagents}, personalized evaluation~\cite{cai2025personalized}, new evaluation frameworks~\cite{kapoor2024aiagentsmatter, yadav2019evalaibetterevaluationsystems, zhuge2024agentasajudgeevaluateagentsagents}, and agent cards~\cite{aiagentsdirectory2025, staufer2025auditcardscontextualizingai}. However, these evaluations share a common limitation: they narrowly focus on technical competence and elide dimensions that require human-interaction such as usability, interpretability, trust, and user satisfaction.
Salaudeen et al.~\cite{salaudeen2025measurementmeaningvaliditycenteredframework} observe that many AI evaluation frameworks suffer from notable validity gaps, and Wallach et al.~\cite{wallach2025position} argue that measurement tasks in generative AI lack the scientific rigor needed for valid conclusions, proposing as redress a framework rooted in social science measurement practices.
In short, evaluating only whether an agent \textit{is capable of} completing a task is insufficient --- we must also evaluate whether users can effectively direct, oversee, and collaborate with those agents to complete tasks \textit{effectively}.
We contribute an evaluation of commercial AI agents with real people, doing tasks representative of how those agents are marketed.

\textbf{Empirical Studies of Human-AI Interaction.}
There is a rich tradition of prior empirical work noting usability breakdowns in human-AI interaction. This work emphasizes that effective human-AI teaming hinges on users’ mental models, expectations, and trust calibration~\cite{ong2012closing, zhang2020ideal, kaur2019sharedmentalmodels, Li02012025}. Furthermore, prior work suggests that framing and communication shape adoption and reliance on AI systems~\cite{khadpe2020metaphors, yin2019understanding, kim2024imnot}, and that explanations should follow human norms (e.g., being selective and conversational) rather than overwhelming users with undifferentiated detail~\cite{Miller2019ExplanationAI}.  Complementary work examines collaboration dynamics and breakdowns: how people interpret competence signals in mixed-initative teams~\cite{10.1145/3491102.3517791}, how system updates can disrupt user experience by shifting mental models~\cite{Bansal2019, nyt2025chatgptcold}, and how specification burdens and communication mismatches complicate collaboration with language-based systems~\cite{zamfirescu-pereira2023johnny, Ko2004SixLearningBarriers, bansal2024challengeshumanagentcommunication}. Emerging studies of agentic LLM systems further highlight the importance of involving users in planning, avoiding over-delegation assumptions, and aligning procedures with user expectations~\cite{he2025planthenexecute, jiang2024fulldelegation, shao2025futurework, wang2025howdoagents}.

Building on this literature, we focus on a distinct empirical gap: how novice end-users experience \emph{operationally agentic} AI agents that \emph{act} on users behalf' by autonomously executing multi-step workflows in real software environments. In these settings, breakdowns can compound across steps and lead to real-world side effects (e.g., committing irreversible actions, producing flawed deliverables, or requiring costly recovery), shifting the core interaction problem from prompting a model to delegating, overseeing, and intervening in ongoing agent activity.

\section{Taxonomy of AI Agents}
We started by taxonomizing the litany of \emph{marketed-as-agentic} products and services being advertised for immediate or near-term use under the ``AI agent'' banner (RQ1). This taxonomy, in turn, helps clarify what the technology industry considers an ``AI agent'' and serves as a \emph{marketing-driven use-case typology}: it captures how products are framed for consumers and what expectations they set. We focus on industry-marketed products because these offerings more clearly define the menu of possibilities for broad consumer use of AI agents --- while there is also much academic interest in the topic, the use-cases explored in academia may be less immediately relevant for end-consumers.

\textbf{Methodology.} We assembled a corpus of $N=102$ \emph{marketed-as-agentic} products by drawing on aggregator lists (e.g., AI Agent Directory~\cite{aiagentsdirectory2025}, Product Hunt~\cite{producthunt}, Google Cloud use cases~\cite{cloud_google_genai_2025}) and keyword searches (``AI agents'', ``LLM agents''). We included products that were marketed or described as ``AI agents'' (or equivalent), were consumer-facing with concrete marketed use-cases, and targeted general end-users rather than only developers or domain specialists. We did \emph{not} require that products fully instantiate the \emph{operationally agentic} properties discussed prior; our goal for RQ1 is to capture how the ``agent'' label is applied in industry, so we coded what capabilities are \emph{claimed} in marketing materials. Overlap and broadness across categories are expected because marketing typologies are not mutually exclusive. Two authors performed an inductive qualitative content analysis~\cite{saldana2021coding} of marketed capabilities, built a codebook through independent coding and consensus~\cite{mcdonald2019reliability}, and abstracted codes into higher-level categories following established taxonomy development practices~\cite{nickerson2013taxonomy}. ``AI agents'' is a rapidly evolving product category; we do not claim an exhaustive census, but a broad sample sufficient to characterize industry-envisioned use cases. Full search and exclusion criteria, analysis details, and the codebook are in Appendix~\ref{ref:appendix-systematic-review-method} and~\ref{ref:appendix-systematic-review-codebook}.

% \subsection{Taxonomy of AI Agents}

\begin{table*}[t]
\centering
\small
\begin{tabular}{@{}p{2.2cm}p{5.4cm}p{4.2cm}@{}}
\toprule
\textbf{Category} & \textbf{Definition} & \textbf{Examples} \\
\midrule
\orchestration{} & Act on behalf of users to manipulate other software (GUIs, automation). & Agentforce; Copilot \\
\creation{} & Generate structured artifacts: slides, apps, docs, marketing materials. & Lovable; Gamma \\
\insight{} & Support analysis, synthesis, recommendations, and information retrieval. & Perplexity Deep Research; Deloitte Care Finder \\
\bottomrule
\end{tabular}
\caption{Summary of three marketed use-case categories for AI agents (full capability map in Appendix~\ref{ref:appendix-taxonomy-full-table}).}
\label{tab:agent-taxonomy-summary}
\end{table*}

\textbf{Findings.} We identified three broad categories of marketed use-cases: \orchestration{}, \creation{}, and \insight{} (Table~\ref{tab:agent-taxonomy-summary}). Categories are not mutually exclusive and, in fact, complex knowledge work often combines insight (problem decomposition), orchestration (executing workflows), and creation (formatting deliverables). The taxonomy thus encodes the \emph{promises} and expectations that \emph{marketed-as-agentic} products set for consumers. We selected one study task per category to test whether those promises are realizable in practice; the usability barriers we report below are, in that sense, expectation failures.

\orchestration{} agents act in GUIs on the user's behalf: they read interface state (e.g., via vision-language models), generate commands with an LLM, and execute them via a controller, producing actions with real-world side effects and raising delegation and oversight demands. Examples include Salesforce Agentforce~\cite{agentforce}, LiveX~\cite{livex}, and Expedia's Trip Matching Agent~\cite{trip_matching_ai_agent}.
\label{ref:taxonomy-orchestration}

\creation{} agents produce structured artifacts (slide decks, websites, apps, documents) with an emphasis on formatting and presentation rather than analytical content; they differ from \insight{} in focusing on deliverable form. Examples include Lovable~\cite{lovable2025} (apps/websites) and Gamma~\cite{gamma} (slide decks).
\label{ref:taxonomy-creation}

\insight{} agents support research, synthesis, and recommendations by combining web search, APIs, and knowledge bases in multi-turn interactions. They address analysis and decision-making that users would otherwise perform manually. Examples that clearly instantiate the operational properties above include Perplexity Deep Research~\cite{perplexity2024deepresearch} and Deloitte Care Finder~\cite{aws_care_finder_agent}. The full capability map with subcategories and counts, including boundary cases (e.g., recommender-style products marketed as agents), is in Appendix~\ref{ref:appendix-taxonomy-full-table}.
\label{ref:taxonomy-insight}
\section{User Study}
We next conducted a usability study to assess whether and how end-users could use \emph{operationally agentic} AI agents for orchestration, insight, and creation tasks in practice (RQ2).

\subsection{Methodology}
We conducted a think-aloud user study with $N=31$ participants, with each 1 hour session comprising two AI agent tasks followed by a semi-structured exit interview. During the study, participants would attempt to accomplish each task using one of two commercial AI agents. These tasks were randomly selected from a set of three we had pre-defined, one for each category of our AI agent taxonomy (i.e., insight, orchestration, creation). The two commercial AI agents were Operator \footnote{\url{https://openai.com/index/introducing-operator/}} and Manus \footnote{\url{https://manus.im/}}.

\subsection{The Participants}
Participants were recruited via social media (X, LinkedIn) and Prolific. The sample spanned ages 18--65+, included 23 men and 8 women, and was largely composed of frequent users of generative AI tools (e.g., chatbots). However, participants were all novices to the multi-step, task-oriented systems marketed as ``AI agents'' that we study here. Full demographics are in Appendix~\ref{ref:appendix-user-study-participants} (Table~\ref{tab:demographics}).

\subsection{The Agents}
We studied two commercially deployed \emph{operationally agentic} products: Manus~\cite{manus2025} and OpenAI Operator~\cite{openai_operator_2025} (now integrated into ChatGPT Agent~\cite{openai2025chatgptagent}). Both support multi-step task execution via a chat-based prompt interface and can use external tools (including computer use) to act in real software environments, with user takeover and oversight; they thus instantiate the operational agentic properties we discussed earlier (i.e., delegation an initiative allocation, acting in an environment on the user's behalf, multi-step execution, and providing mechanism for oversight, intervention, and recovery). We selected them because they were publicly accessible and covered all three taxonomy categories at the time of study. An extended description of both agents and their interfaces is in Appendix~\ref{ref:appendix-agent-interfaces}.
 
\paragraph{Manus.}
\begin{figure}[ht]
    \centering
    \includegraphics[width=\linewidth]{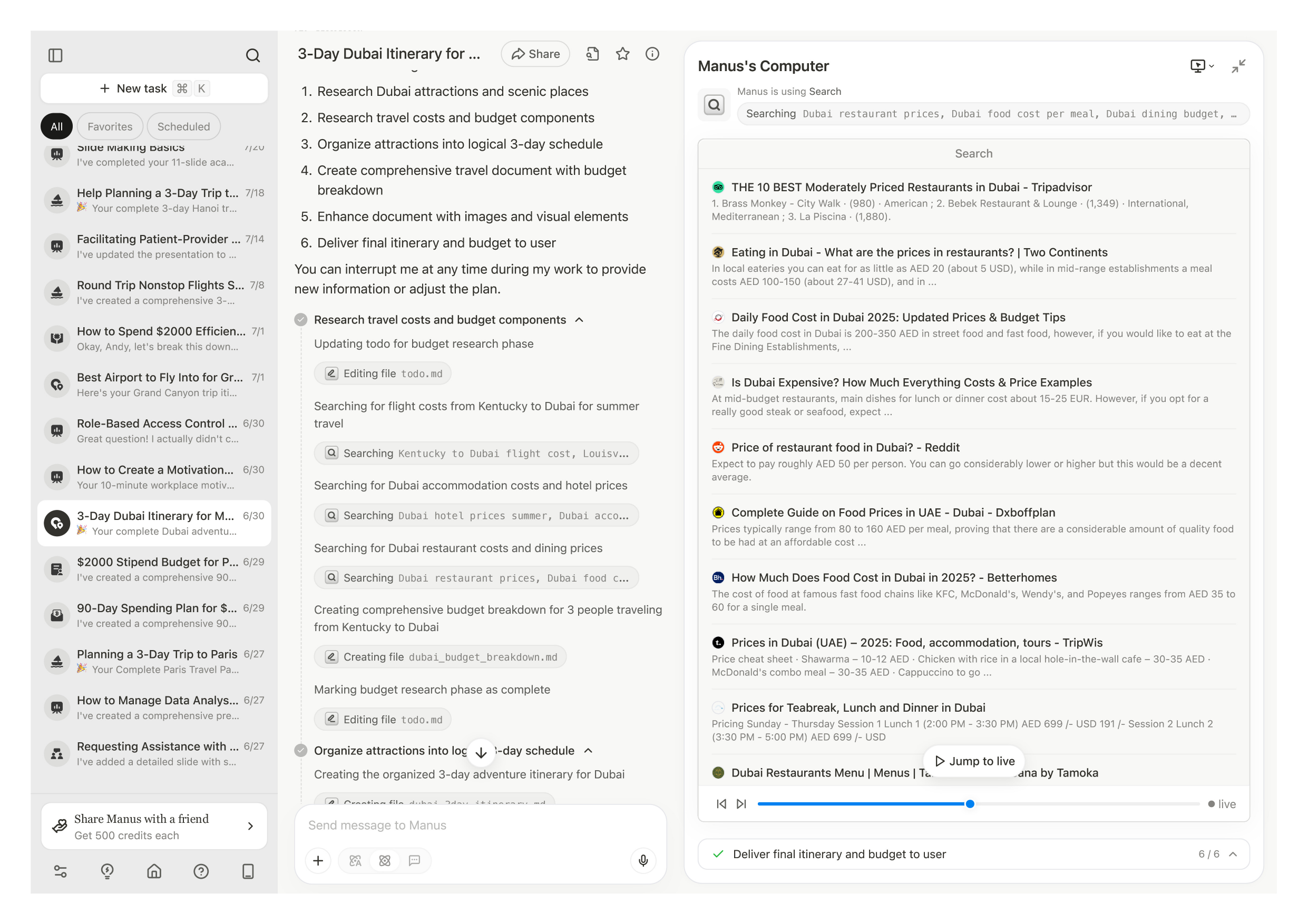}
    \caption{Manus, one of our AI Agents, working on our Holiday Planning task.}
    \label{fig:manus-intro}
\end{figure}
Manus is a general-purpose AI agent that can perform web searches and control a computer (browser, terminal, filesystem) from a persistent workspace UI (Figure~\ref{fig:manus-intro}). Its interface combines a chat thread, live computer view, and progress indicators, making ongoing work visible. When a user sends a prompt, Manus typically outlines steps and emits a detailed action log with links to intermediate states and outputs. Users can also intervene mid-task via ``Take Over,'' manipulate the computer, and return control.

\paragraph{Operator.}
\begin{figure}[ht]
    \centering
    \includegraphics[width=1\linewidth]{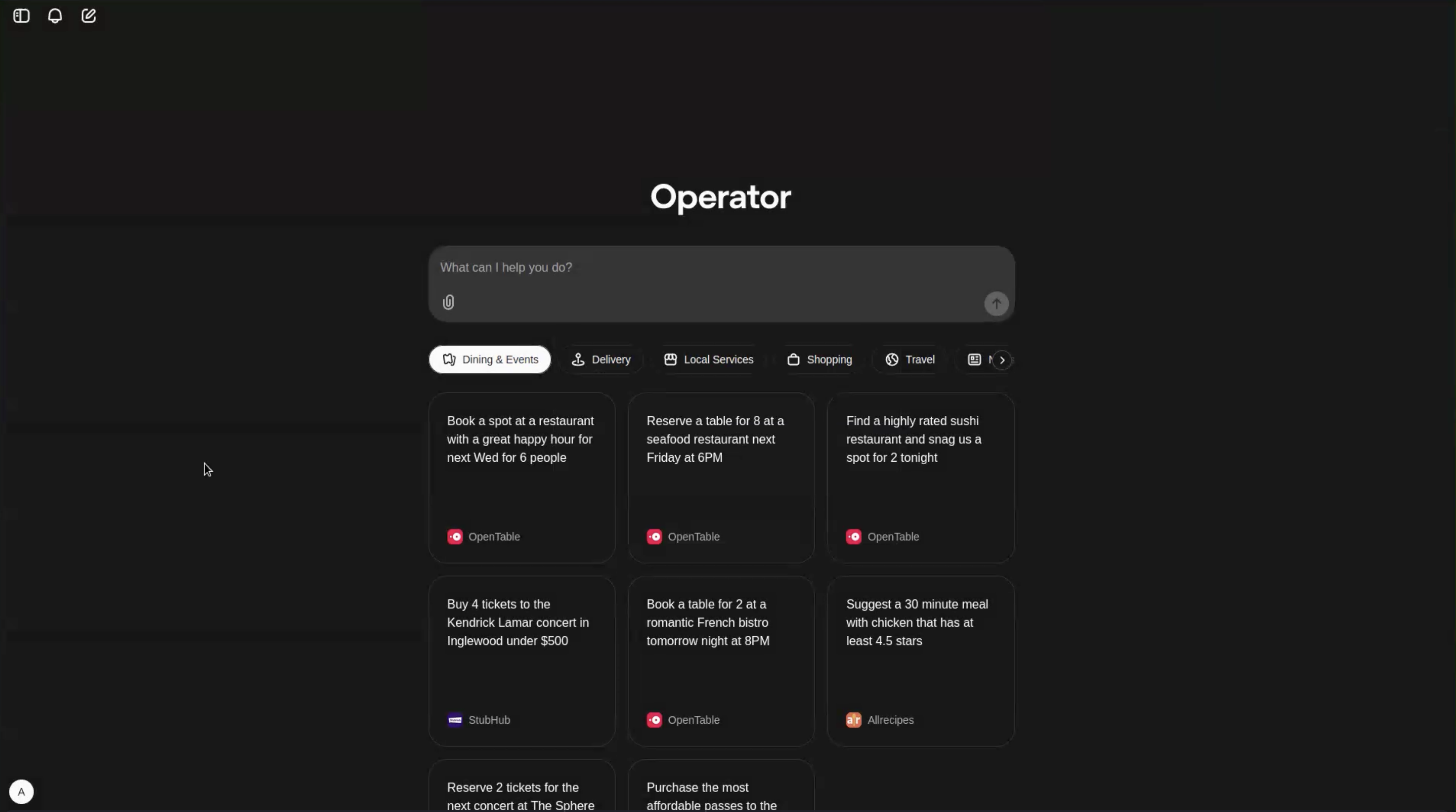}
    \caption{The initial screen on Operator, featuring a text box to enter prompts, and a grid of example tasks in various categories.}
    \label{fig:operator-intro}
\end{figure}
Operator offers comparable computer-using capabilities but with a more minimal, chat-forward interface (Figure~\ref{fig:operator-intro}) and non-streaming responses. It provides task idea cards on entry and then transitions to a standard chat view after the first prompt. When it invokes computer use, it embeds a computer preview in the chat thread (expandable to a larger layout) and supports user takeover for required steps (e.g., authentication) before returning control to the agent. 

\subsection{The Tasks}
We defined three tasks for our participants to attempt during the study --- one for each category of our AI agent use-case taxonomy. We designed this study without a non-agent baseline condition because our objective was to explore the novel interaction challenges that arise when end-users delegate tasks to autonomous agents, not to compare agent performance against manual task completion. Tasks were approximately evenly distributed across participants and agents (see Table~\ref{tab:sus_stats}). Specifying tasks is, of course, a very open-ended activity. To pick useful tasks for our study, we selected tasks that were:
(i) \textbf{familiar} to non-specialists, (ii) \textbf{feasible} within the 20-minute think-aloud window, and (iii) \textbf{emotionally salient} enough to elicit meaningful preferences and evaluation. Full task-selection criteria and task screenshots are in Appendix~\ref{ref:appendix-study-task-design}.

\paragraph{\orchestration{} Task: Holiday Planning.}
In this task, participants used an AI Agent to produce an itinerary of a 3-day holiday to a location of their choice, including flight tickets, housing arrangements, sightseeing, and other activities of interest. This task involved orchestration (Section ~\ref{ref:taxonomy-orchestration}) in that it requires the agent to complete individual steps by navigating real UIs on an active computer. The task required the agent to navigate flight aggregators, elicit user preferences for activities, and synthesize results into a coherent schedule under complex constraints (budget, seasonal availability, personal preferences). Holiday planning is a familiar, emotionally salient task --- the non-trivial costs and personal nature of travel raise the stakes of delegation.

\paragraph{\creation{} Task: Slide Making.}
In this task, participants used an AI agent to create (Section ~\ref{ref:taxonomy-creation}) a slide deck for a 10-minute presentation on a topic of their choice. We instructed users that the slides should be something they would be willing to present themselves, raising the stakes and requiring critical evaluation of outputs. Preparing presentations is common in knowledge work, and people have strong variation in both stylistic and content preferences, making this a task that naturally requires collaboration between users and agents.

\paragraph{\insight{} Task: Professional / Personal Growth Stipend Budgeting.}
In this task, participants used an AI agent to help them make the best use of USD 2,000 to advance their personal or professional development. This task required insight (Section ~\ref{ref:taxonomy-insight}) in that it the agent had to understand user goals, perform web searches, evaluate options against budget constraints, and synthesize results. Budgeting is a familiar, emotionally salient task --- what each person considers a good use of disposable money is highly dependent on their values and goals, making a generic output unlikely to satisfy.

\subsection{Procedure}
We ran 60-minute Zoom sessions with $N=31$ participants. Our sample size is consistent with qualitative study benchmarks: for example, Caine’s analysis of CHI papers reported a modal sample size of 12 \cite{caine2016local}, and other highly cited studies on sample sizes for qualitative research found that saturation is commonly reached within 12-24 interviews \cite{guest2006many,hennink2017code}. Each participant completed 2 assigned tasks using Manus or Operator, each of which included a 20 minute think-aloud session and a semi-structured exit interview. We collected screen and audio recordings, System Usability Scale~\cite{brooke1996sus} responses, and interview transcripts.
% We used reflexive thematic analysis with iterative consensus coding by two authors.
Tasks and agent tools were counterbalanced, with combinations approximately evenly distributed across participants (see Table ~\ref{tab:sus_stats}).
The interviewer did not answer questions about the agent interface except in limited circumstances where the tool appeared to be broken. Our recruitment flow, compensation, assignment details, and protocol are in Appendix ~\ref{ref:appendix-study-logistics} and our interview script is in Appendix ~\ref{ref:appendix-user-study-script}. The study was approved by an institutional review board (IRB).

Consistent with the think-aloud protocol, the interviewer periodically asked participants to articulate what they thought was happening, what they expected to happen next, and how they evaluated the agent's progress. These organic ``sync'' moments became especially important when progress halted due to apparent platform or tool failures: the interviewer probed participants' interpretations of the blockage and how they were thinking about recovery, but did not provide task instructions. For example, some participants treated failures as technical issues on their own end, some evaluated the agent favorably despite the blockage, and some attempted to pivot to tools or workflows with fewer login requirements. We did not code sessions differently by intervention status; instead, these moments were analyzed as naturally occurring windows into participants' mental models of agent failure modes, intervention possibilities, and recovery strategies.

To analyze the data we collected, we conducted a reflexive thematic analysis~\cite{braun2006thematic} to understand what barriers and usability challenges end-users faced while attempting the tasks we had assigned them to complete with the use of an AI agent (RQ2). The first and second authors independently coded three interviews using open inductive coding~\cite{saldana2021coding}, generating an initial set of codes grounded in concrete, low-level challenges participants encountered. They met weekly and discussed the codes for multiple interviews, merging and de-duplicating codes when realizing they were either the same phenomenon or variants that were different experiences altogether. Through this iterative process, the authors abstracted recurring patterns into higher-level themes that explained general experiences participants went through as they navigated agent software. Consistent with best practices for qualitative research in the HCI community~\cite{mcdonald2019reliability}, agreement was reached through discussion and consensus rather than inter-rater reliability metrics. Our codebook can be viewed in Appendix ~\ref{ref:appendix-user-study-codebook}.

\subsection{Findings}
\begin{figure}[ht]
    \centering
    \includegraphics[width=\linewidth]{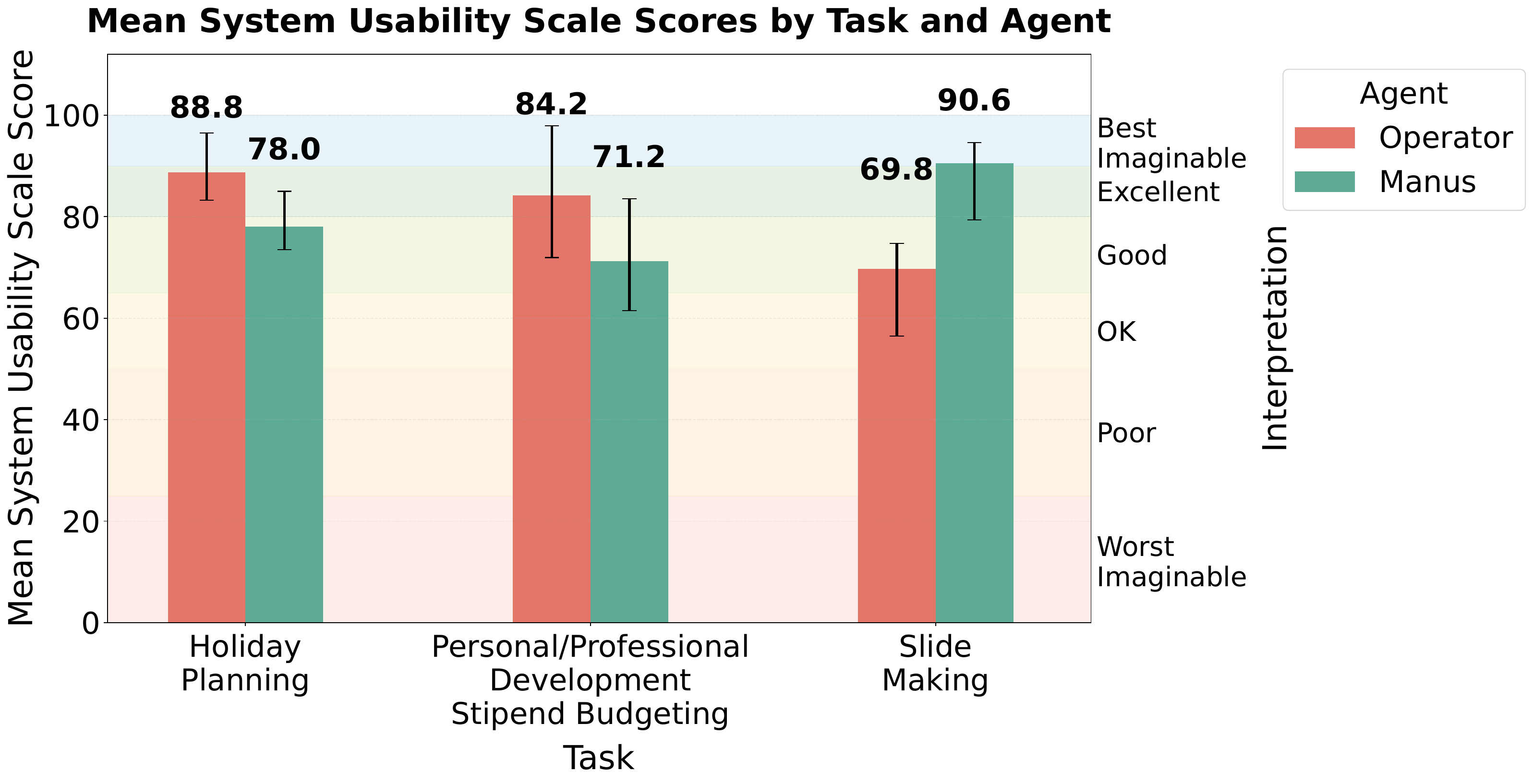}
    \caption{Mean System Usability Scale (SUS)~\cite{brooke1996sus} scores by task and agent. Error bars show 95\% bootstrap confidence intervals (10{,}000 resamples); per-cell sample sizes are $n=9$--$13$. Per-task standard deviations and Welch's $t$-test results are reported in Table~\ref{tab:sus_stats}.}
    \label{fig:sus_score_plot}
\end{figure}

Participants were generally able to complete both assigned tasks with a reasonable degree of success --- though we specifically selected tasks that we verified were simple and doable prior to the study. Nevertheless, we identified a number of critical usability barriers that hindered participants' ability to make optimal use of the AI agents we tested. We will start by discussing where the agents succeeded, general impressions towards the agents, and then discuss the usability barriers we identified.

\subsubsection{Agent successes}
Figure~\ref{fig:sus_score_plot} and Table~\ref{tab:sus_stats} report SUS scores~\cite{brooke1996sus} by task and agent. We compare agents within each task using Welch's two-sample $t$-test, which does not require equal variances between groups. Seven participants completed the same task with both agents, partially violating independence; we treat these tests as descriptive context rather than strict inference. Manus rated significantly lower than Operator on Holiday Planning ($-10.8$ points, $p=.043$) and significantly higher on Slide Making ($+20.8$ points, $p=.012$); Stipend Budgeting was not distinguishable ($-12.9$ points, $p=.184$). With $n=9$--$13$ per cell, the study is underpowered to detect modest effects, and contrasts are unadjusted for family-wise error.

\begin{table*}[t]
\centering
\small
\caption{Per-task SUS scores by agent. The Manus$-$Operator estimate, 95\% CI, and $p$ value are from Welch's two-sample $t$-test (which does not assume equal variances between agents). Cells show mean ($M$) and standard deviation ($SD$).}
\label{tab:sus_stats}
\begin{tabular}{lcccccc}
\toprule
Task & \multicolumn{2}{c}{Operator} & \multicolumn{2}{c}{Manus} & Manus$-$Operator & \\
\cmidrule(lr){2-3} \cmidrule(lr){4-5}
 & $n$ & $M$ ($SD$) & $n$ & $M$ ($SD$) & estimate [95\% CI] & $p$ \\
\midrule
Holiday Planning & 10 & 88.8 (8.6) & 10 & 78.0 (12.8) & -10.8 [-21.1, -0.4] & $0.043$ \\
Stipend Budgeting & 9 & 84.2 (18.2) & 10 & 71.2 (22.4) & -12.9 [-32.6, +6.8] & $0.184$ \\
Slide Making & 10 & 69.8 (20.9) & 13 & 90.6 (8.6) & +20.8 [+5.4, +36.2] & $0.012$ \\
\bottomrule
\end{tabular}
\end{table*}

Participants were charitable toward both agents even when tasks failed: \textit{``This is worth it in every sense of the word''} (P23); most praised the agents' detail (\textit{``very comprehensive and thorough''}, P26). None had used AI agents before, so this may partly reflect novelty. The favorable SUS scores may appear in tension with the barriers below; we find both genuine. Participants recognized real value, but the barriers---around mental models, trust, collaboration, communication, and metacognition---are frictions that would compound with sustained use, and would likely be magnified in higher-stakes, less structured real-world settings beyond our deliberately simple tasks.

Participants also reported a consistent tradeoff: Manus produced richer first-class outputs (especially for slides) but introduced higher communication overhead and lower editability of intermediate artifacts, while Operator was more concise but stalled more often during computer-use workflows. Both exhibited the barrier classes below.

\subsubsection{Barrier 1: Agent behaviors misalign with user mental models.}
\label{ref:barrier-1-alignment}
Many participants were unsure what agents could do, how much detail prompts required, and what role they should play during execution. This uncertainty made initial prompting feel high-stakes and encouraged trial-and-error interaction rather than confident delegation. As one participant noted, \textit{``You've got to sit there and make sure that your initial prompt is perfect \ldots You ask exactly what you're looking for.''} (P26).

Users also misunderstood core interaction mechanics (e.g., when and why takeover was needed), suggesting that current interfaces under-specify agent affordances and execution boundaries. Overall, users built mental models reactively from outcomes rather than proactively from clear system cues, contributing to downstream trust and control breakdowns.

\paragraph{Users were apprehensive and uncertain about prompting agents.}
End-users generally did not know what to expect when sending an agent a prompt, leading to a behavior that we term \emph{``prompt gambling''}, i.e., users expressing uncertainty and hesitation with prompting agents owing to concerns that it would generate unnecessary and excessive outcomes.

This uncertainty led to a range of different prompting strategies. Some users believed they needed to decompose a task into individual steps, and have an agent perform each step one at a time. These users would craft an initial prompt with only the first step of the task --- but the agents they used would assume that this first step was the full task, leaving users feeling trapped. Other users would provide a highly detailed initial prompt to try and prevent agents from going astray. The ultimate effect was that many participants felt that the initial prompt they fed to the agent was high-stakes and essential to get right.

In other cases, users were pleasantly surprised by how much the agent could do with little specification: \textit{``I thought I needed to give it source material for the slides, but it did all the research and organized it in slides that would've taken me weeks to put together.''} (P26) These participants gradually updated their model of the agent from a deterministic tool requiring structured hand-holding to something more \textit{``as if it were an administrative assistant''} (P9). This confusion --- while not stifling --- still indicates that agents' \textit{affordances} were mysterious to users; their conceptions arose only after trial-and-error, and even then remained task-specific and incomplete.

\paragraph{Users had trouble understanding what agents did, and why.}
Multiple users were confused about what ``Computer use'' was and how it worked. Attempting to make slides, Operator repeatedly asked users to use \emph{Take Over} to sign in to Google Slides, which felt unexpected and undesirable. One user expected the slide outline and designs before being asked to log in, and expected slides to be prepared within Operator, unaware that slide-making was not native functionality. They also misunderstood \emph{Take Over}, believing it was a request for them to make the slides rather than sign in and return control to the agent. P19 stated: \textit{``it’s kind of like I’m giving you a job, and you’re throwing the job back at me...You have not even given me anything yet, and you want me to do some stuff for you already.''}

A more technically knowledgeable participant questioned the need for Computer Use altogether, preferring web APIs for research tasks over browser control. Computer Use was perceived as far slower than participants' own manual searches.

Some participants developed folk models of how interim outputs affected speed. For example, noting the volume of progress reports, P31 wondered whether hiding workflow logs would make the~agent~faster.

In short, AI agents are capable enough that simple recall-based chat leaves too much to users' imagination. This creates confusion about what agents can do, what they need from users, how they work, and what role users play in task completion.

\subsubsection{Barrier 2: Agents presume trust in sensitive contexts.}
\label{ref:barrier-2-presumption-of-trust}
Several participants (e.g., P22, P26, P30) expressed distrust in working with AI agents to perform either specific parts of a task or the task as a whole. Trust issues with each of the tools spanned concerns around hallucinations, password management, low expectations of agents' capabilities, and disinclination to hand over too much autonomy over tasks users preferred to handle personally.

Not all users wanted to trust the agent with credentials to their accounts: \textit{``I don’t particularly want to give it my Google account information''} (P30).

Some users also expressed apprehension about entrusting the agent to do tasks on their behalf, such as planning a trip:  \textit{``I would plan manually first, then compare… I still don’t trust it completely''} (P26). P22 further explained how these trust issues were compounded by the fact that the agent did not ask them questions about their preferences or constraints: \textit{``it didn't ask me. you know. Do I want one bed or 2? Do I want a a room that looks out on the pool? Or do I want one that looks out on the plaza?''} This example highlights how trust-building is discursive; for tasks that require capturing personal preferences, overly eager agents that run off and do work without confirming user intent can reduce trust.

Several participants also expressed concern about the veracity of the information agents provided, especially when the values conflicted with \textit{a priori} expectations or seemed too good to be true.

\subsubsection{Barrier 3: Agents fail to accommodate diverse collaboration styles.}
\label{ref:barrier-3-diversity-of-collaboration-styles}
Many participants were uncertain how to ``collaborate'' with agents; several requested pause/stop or finer-grained control (e.g., P23, P26). Users were unsure what they could delegate, how to exert control, and how to recover from errors.

\paragraph{Users lack effective control mechanisms for steering agent behavior during task execution.}

When agents failed or otherwise produced poor initial outputs, several users believed that it would be faster and more effective to restart from scratch as opposed to asking the agent to iterate on the provided deliverable.
They did not believe the agent could effectively make fine-grained edits. P31 captured this feeling when expressing confusion as to why the agent was able to easily create new slides, but had trouble following more detailed instructions to iterate on these slides:

\begin{quote}
    \textit{``I remember the original one. I mean, it had, like maybe one more block of text on the left hand side, and just for it to remove, you know, that seemed like it took longer than you know, creating all 7 other slides. So that was kind of I don't know why it got so caught up on that.''} --P31
\end{quote}

Users also expressed wanting better real-time control mechanisms to supervise the agent and ensure it stayed on track. Participant 26 expressed this tritely, \textit{``I’d like a pause button if it’s going off the rails.''} Similarly, P23 expressed wanting a ``stop button'' to exert more control over how long the agent worked per prompt.

Others missed or were hesitant to oblige Manus' message inviting additional prompts even while it was responding to a previous prompt. These participants questioned if sending a new prompt would derail the agent or otherwise cause it to lose progress.

\paragraph{Users vary in the level of proactivity they would like out of agents}

We observed strong variation in how involved individual users wanted to be with crafting the final output delivered by the agent. For example, for the slide making task, some users were interested in being deeply involved with the design of each slide, whereas others were content with the agent proposing its own outline and building a slide deck around it, with little involvement from them.
% P26 shared: ``\textit{I would plan manually first, then compare… I still don’t trust it completely}".
P9, for example, mentioned ``\textit{my typical way of doing this kind of thing is to build all this as independent units on my own.}"

More generally, we found that some users viewed agents as ``thought partners'' while others viewed agents as execution tools.  P16 was emblematic of the former camp: \textit{``I like to do some of the baseline stuff myself… use AI more as… confirmation...making sure that what I'm taking away from something is maybe correct…and then kind of just verifying that I didn't miss anything''} Agents generally did not appear to account for this variation, and tended towards being over-eager execution tools --- i.e., assuming users wanted minimal overall involvement. 

\paragraph{Agents made incorrect assumptions about users, leading to miscalibrated outputs}

Agents also appeared to sometimes make incorrect assumptions about users which could lead to deliverables that were misaligned with user expectations.

For example, for the Slide Making task, P30 requested that the agent make a slide deck about bats. The final slide deck the agent produced was minimal --- just a stream of bat images with little text to scaffold the rationale for each image. This lack of scaffolding left P30 feeling frustrated, and feeling like the agent made a faulty assumption: \textit{``It thinks I'm an expert at bats, but I don't know much, and I certainly can't present a talk with slides that just have photos''}.
P30 further expressed wanting to explicitly specify roles and expectations in the collaboration: \textit{``Hey, I'm really good at doing this. These are things that you might need. I might need your help on \ldots.}''

Similarly, for the Trip Planning task, P13 found many of the travel suggestions the agent made to be fairly basic, and only necessary for people taking a flight or visiting a country for the first time, an assumption that they believe the agent had made incorrectly.

\paragraph{Users struggle with recovering from agent errors due to opaque failure modes.}
\begin{figure}[ht]
    \centering
    \includegraphics[width=1\linewidth]{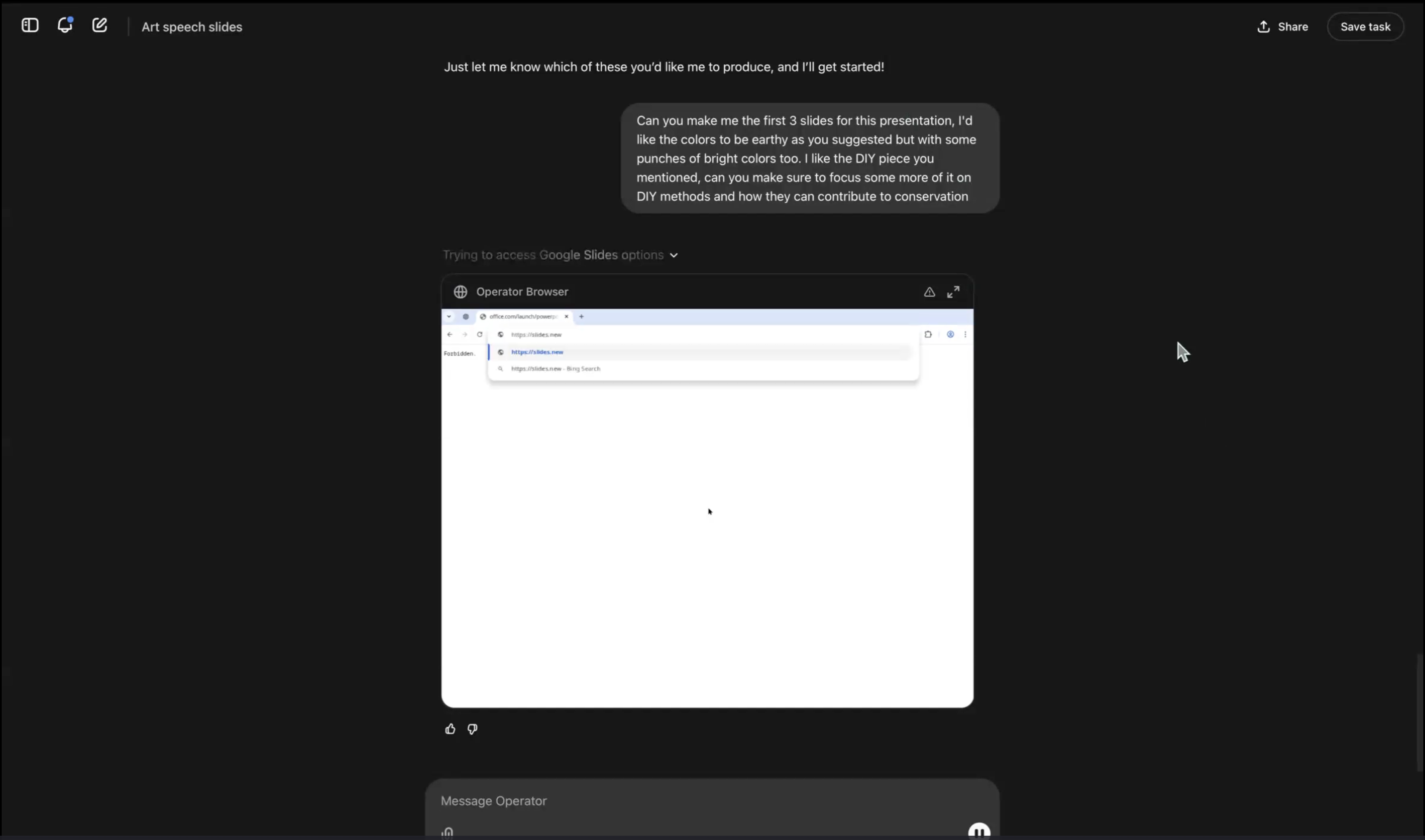}
    \caption{OpenAI Operator, working on the slide making task. Frequently, it was blocked while connecting to online presentation software, potentially due to anti-bot protection.}
    \label{fig:Operator-Computer-Use}
\end{figure}

Agents occasionally made errors as they attempted to complete tasks --- indeed, 14 out of 31 users encountered at least one operational error during the study, such as being blocked by bot detection workflows (~\ref{fig:Operator-Computer-Use}). Many users struggled with recovering from these errors.

In particular, for Operator's Computer Use, many users did not even register errors when they did occur.
Operator generally froze in these circumstances with no error message or troubleshooting instructions, leaving users uncertain: \textit{``I wasn't quite sure why it was failing.}'' (P7)
As a result, many users attributed Operator's lack of progress to arbitrary causes such as websites being incompatible with Operator, security settings on the user's computer, or because of their own errors. P30, for example, said: \textit{``I didn't realize it was failing... that could just be my own ignorance of the platform~it~was~using.''}

These complete blockages prompted interviewer probes about participants' interpretations of the failure and recovery options, but not direct task guidance.

\subsubsection{Barrier 4: Agents create communication overhead for users.}
\label{ref:barrier-4-overwhelming-communication-overhead}
Participants across the spectrum of progress-visibility preferences (e.g., P16, P21--P24, P31) reported difficulty parsing agent output or articulating intent.

\paragraph{Users have varying expectations around communication of agent progress.}

We observed a full spectrum of expectations and preferences around how agents should communicate task progress. Some (e.g., P16, P24, P31) preferred final-output-only interaction, while others (e.g., P21, P22, P23) wanted progress updates for oversight and confidence.
For example,
% P31 shared: \textit{``I just want the thing done, but not necessarily know how it's done.''}.
% On the other hand,
P16 stated: \textit{``It’s just too much to take...Oh, my God! It threw out so much stuff...it’s almost an overwhelming amount of information''}. Across both preferences, users reported that the logs were hard to parse because key milestones were buried in dense, uniformly formatted output.

\paragraph{Prompting agents is a technical specification recall task that is cognitively demanding.}

Many participants expressed that prompting agents was difficult, because they were themselves uncertain about how much they needed to communicate with the agent and how best to communicate the web of interconnected preferences they had in their minds. 
\begin{quote}
    \textit{something that might be obvious to a person, is maybe not always obvious to the AI is needed information. So it doesn't think to ask, and you don't think to tell it and then realize there's a gap.} --P16
\end{quote}
For instance, for the Slide Making task, users expressed a desire to directly modify slides agents had initially improved, rather than asking the agent to iterate on those slides. These users believed it would be too difficult to articulate, in a way agents might be able to understand, their criteria for what they wanted to improve. They simply wanted ``better images'' or ``less text-heavy'' slides.

Similarly, on the Trip Planning task, many people wanted to pick flights themselves because of complex, conditionally intertwined criteria that were difficult to express as structured instructions (e.g., frequent flyer miles, number of stops, time of day). The perceived need to do so created intense metacognitive demands.

\subsubsection{Barrier 5: Agents lack metacognitive self-assessment.}
\label{ref:barrier-7-assess-and-deliver-quality-outcomes}
Several users (e.g., P16, P19, P20, P29) observed that agents struggled to assess their own progress, limitations, and output quality in terms users could understand.

\paragraph{Agents don't know what they don't know and can't do.}
When agents encountered difficulties or made errors, several users observed that agents would fail to recognize these errors and get stuck in a repetitive loop. For example, for the Slide Making task, Operator 
didn't seem to recognize its inability to control some slide-making websites, which remained frozen for extended periods of time. 
P16 noted: \textit{``It doesn't have access to certain things, so it couldn't do it. And it just was kind of circling.''}
P16 then lamented that if they had taken more initiative in troubleshooting, they would have been able to compensate for the agent's failure to recognize its limitations, highlighting meta-cognitive gaps that users must sometimes cover.

\paragraph{Users expressed a desire for agents to express greater reflective abilities.}
For example, for the Personal Growth Insight task, P20 noticed that Manus hallucinated information about courses that didn't exist at all, even when synthesizing information from Web searches. In this situation, the user (P20) preferred a different response: \textit{``it's seeking to provide an answer rather than to say, I don't know, or to link someone where they should go to find information.''}
Similarly, for the Slide Making task, P19 expected the agent to take more active role in seeking critique to improve the design: \textit{``It probably covered like 70 to 80\% of what I was expecting. I was waiting for it to ask me for some more information.''}

\paragraph{Agents use tools with which users are unfamiliar, precluding effective oversight.}
We observed that agents sometimes used tools with which users were unfamiliar, inhibiting oversight and control.
We often saw this in the Slide Making task: agents unexpectedly used programming tools to create slides. P29, for example, was surprised by Manus' decision to use HTML: \textit{``I wasn't expecting it to make the slides all in HTML!''}
When Operator did something similar, P25 felt left out of the process and pessimistic about success.

\section{Discussion}
% \subsection{From Industry Framing to User Reality}
% \begin{figure*}
%     \centering
%     \includegraphics[width=1\linewidth]{figures/Usability-Barriers-to-Design-Implications.pdf}
%     \caption{Mapping five empirically discovered usability barriers to six implications for next-generation AI agent design.}
%     \label{fig:usability_barriers_mapped_to_design_implications}
% \end{figure*}
Our results show a gap between industry aspirations for AI agents and how novice users can reliably use them in practice. In taxonomizing the capabilities of 102 products marketed as ``AI agents'' and conducting a think-aloud usability study with 31 participants assessing these capabilities across two operationally agentic systems, we found five critical usability barriers: (1) mental-model misalignment, (2) premature trust assumptions, (3) collaboration-style mismatch, (4) communication overload, and (5) weak metacognitive behavior.
% These issues extend beyond interaction: they degrade oversight, controllability, and recovery in AI agent systems during multi-step task execution.
Table~\ref{tab:barriers-to-agentic-properties} maps each barrier to an implicated agentic property.
% This mapping operationalizes the properties that distinguish ``agents'' from standard LLMs that we discussed in the introduction: mixed-initiative and interface agents~\cite{principles_of_mixed_initiative_user_interfaces, shneiderman_maes_1997, maes1994, Lieberman1997Autonomous}; supervisory control and human-in-the-loop agentic systems~\cite{sheridan1992telerobotics, mozannar2025magenticui}; and reason--act/tool-use architectures~\cite{yao2023react, schick2023toolformer}.
\begin{table}[t]
\centering
\small
\begin{tabular}{@{}ll@{}}
\toprule
\textbf{Usability barrier} & \textbf{Implicated agentic property} \\
\midrule
Mental-model misalignment & \makecell[l]{Hidden plans/boundaries \\ Long-horizon autonomy} \\
\hline
Premature trust assumptions & \makecell[l]{Credential delegation \\ Real-world actions} \\
\hline
Collaboration-style mismatch & \makecell[l]{Mixed-initiative planning \\ Intervention points} \\
\hline
Communication overload & \makecell[l]{Step logs \\ State over time} \\
\hline
Weak metacognitive behavior & \makecell[l]{Stall/loop detection \\ Handoff to users} \\
\bottomrule
\end{tabular}
\vspace{3mm}
\caption{Mapping of the five empirically identified usability barriers to the agentic properties (from prior HCI and ML literatures) that cause or amplify them.}
\label{tab:barriers-to-agentic-properties}
\end{table}

\subsection{Relationship to LLM Usability Barriers}
We did not include a chatbot baseline, so our claim is not that these barriers are absent from chatbots. Rather, our results show how familiar LLM usability problems are reconfigured when language-based interaction becomes a mechanism for delegating action. In chatbots, users primarily evaluate and revise model outputs. In operationally agentic systems, users delegate work that unfolds over time through plans, tool calls, external interfaces, credentials, and intermediate artifacts. The usability surface expands --- prompting remains a challenge, but in addition there are questions around if users can appropriately bound an agent's autonomy, inspect its progress, intervene at the right moment, and recover from failures.

Prior work shows that non-experts prompt LLMs opportunistically and struggle to form stable mental models of language-based systems~\cite{zamfirescu-pereira2023johnny, Ko2004SixLearningBarriers, bansal2024challengeshumanagentcommunication}; generative-AI design work similarly emphasizes the difficulty of specifying, prototyping, and controlling variable model outputs~\cite{subramonyam2025prototyping, weisz2024design}.  Our findings suggest that these known challenges become more consequential when ambiguous instructions are enacted rather than merely answered. In a chatbot, a poor prompt may cost another turn. In an agent, the same ambiguity can launch a time- and resource-consuming multi-step execution process that invokes external tools and/or produces irreversible side effects before the user can diagnose the mismatch. Thus, Barrier 1 and Barrier 2 are not only about users knowing what to ask or whether to trust an answer; they are about whether users understand what the agent will do after being asked, where it will act, what assumptions it will make, and what risks it may incur on the user's behalf.

The remaining barriers similarly reflect the shift from response generation to enacted work. \emph{Task diversity} becomes more challenging because agents may manipulate heterogeneous external substrates such as browsers, files, accounts, slide editors, calendars, and software tools in ways that users may not be able to effectively oversee (e.g., when agents create slides programmatically). \emph{Metacognition} becomes central because agents can stall, loop, or use unfamiliar tools over many steps; users need to know when the agent is struggling, what it is doing, and when they should intervene. Finally, \emph{collaboration} differs from turn-based chat because agents may take many actions per instruction while users interleave corrections, clarifications, and new goals.
% These barriers are therefore not merely transient affordance confusion or weak model capability. Better interfaces and models may mitigate them, but delegation, oversight, and recovery remain inherent to agentic systems.

Some manifestations of these barriers may diminish as users gain experience and models improve, but delegation, oversight, intervention, and recovery are structural requirements of agentic systems that will not be addressed simply by expecting more capable users and models in the future. We next translate these barriers into concrete implications for designing and evaluating agentic systems for general end-users.

\subsection{Implications for \emph{Designing} AI Agents}
Our findings yield six high-priority design implications for end-user AI agents,
% (Figure ~\ref{fig:usability_barriers_mapped_to_design_implications}),
each directly grounded in the five usability barriers we observed.

\subsubsection{\textit{``Know your user''}: Calibrate to user preferences}
Users vary in desired proactivity, communication verbosity, and trust requirements (Barriers 2--4). Agents should maintain lightweight user profiles (e.g., preferred collaboration style, verbosity, trusted sources) and ask task-specific preference questions at session start, including acceptable autonomy and when to defer.

\subsubsection{\textit{``Know yourself''}: Improve self-assessment}
Participants encountered agents that missed blocked progress, presented imperfect drafts as final, or used unexpected tools (Barrier 5). Agents should expose confidence, detect repeated failures, and hand off control when stalled or in need of user input. They should also distinguish completed, blocked, and uncertain states so they neither stop too early nor gratuitously revise work that is already acceptable.

\subsubsection{\textit{``Be adaptable''}: Adapt to task and end-user}

Users wanted different levels and formats of progress visibility, but interfaces presented uniform, high-volume logs (Barrier 4). Agents should adapt presentation by task type and user preference, using progressive disclosure to emphasize milestones while collapsing~routine~actions.

\subsubsection{\textit{``Measure twice, cut once''}: Support planning and execution control}
Users lacked control over plan quality after the first prompt and often felt trapped in poor-quality execution paths (Barriers 1 and 3). Agents should support pre-execution plan review/edit and in-execution controls such as pause, stop, rollback, and checkpointing. This is especially important when agent actions consume time, credentials, or real-world opportunities before inspection.

\subsubsection{\textit{``Show, don't tell''}: Support non-textual input}

Purely free-text controls imposed a high recall burden and made nuanced intent difficult to express (Barrier 4, leading to Barrier 1). Agents should combine instruction templates, recognition-based controls, and demonstration-based input~\cite{programming_by_demonstration} so users can specify high-impact choices without relying on prompt-engineering know-how, especially for visual edits and conditional constraints.

\subsubsection{\textit{``Next time's the charm!''}: Support precise iteration}
Participants often trusted agents for first drafts, but not for precise iterations, especially on visual and structural deliverables (Barrier 3). Agents should support artifact-native iteration: inline comments, visual diffs, section reordering, version history, and selective revert. Intermediate outputs should remain inspectable and editable.

\subsection{Implications for \emph{Evaluating} AI Agents}
Based on our findings, we propose that agentic systems be assessed along at least the following axes to ensure usability: (i) \emph{intervention frequency}---how often users must pause, correct, or redirect the agent (Barriers 1, 3); (ii) \emph{time-to-recovery}---latency and effort to recover from agent errors or stalls (Barriers 3, 5); (iii) \emph{stall/loop rate}---incidents where the agent fails to recognize blocked progress or repeats unproductive actions (Barrier 5); (iv) \emph{plan-edit friction}---ease of reviewing or revising the agent's plan before or during execution (Barriers 1, 3); (v) \emph{credential and handoff points}---clarity and safety of sensitive action delegation (Barrier 2); (vi) \emph{iteration utility}---the ability to make specific, targeted improvements on an interim output (Barrier 3); and, (vii) \emph{rollback/checkpoint use}---availability and usability of undo or checkpointing (Barrier 3). These dimensions complement existing capability-oriented evaluations and support the design of more reliable agentic systems.

\subsection{Limitations}
As with any empirical study of an emerging technology, our study has limitations related to the tasks we developed, the composition of our participant pool, and the dynamism of the product space. Our user study features three sample tasks, which, while representative of the use cases from our taxonomy, cannot cover the breadth and depth of tasks that AI agents purport to support and that can be found in modern digital knowledge work. Next, our group of participants came exclusively from the US sociocultural context, were relatively young, able, and comprised frequent users of generative AI (but not agentic AI) tools. Finally, we assessed desktop versions of just two commercial AI agents from a large and rapidly expanding space of agent software, whose models, capabilities, interaction paradigms, and resulting usability challenges may differ.

\section{Conclusion}
In this study, we created a taxonomy of industry-marketed use cases of ``AI Agents'' and conducted a usability study of two commercial agents, Operator and Manus, to examine how marketed capabilities translate in practice. We identified three broad use cases---orchestration, creation, and insight---and observed five recurring usability barriers: (i) mental-model misalignment, (ii) premature trust assumptions, (iii) collaboration style mismatches, (iv) communication overload, and (v) weak metacognitive behavior. We translate these into implications for designing and evaluating end-user AI agents: calibrate to user preferences, support self-assessment, adapt to task and user, broaden intent input beyond free text, and support precise iteration. Together, these identify central challenges to realizing the longstanding aspirational vision of AI agents.

\begin{CCSXML}
<ccs2012>
   <concept>
       <concept_id>10003120.10003123.10011759</concept_id>
       <concept_desc>Human-centered computing~Empirical studies in interaction design</concept_desc>
       <concept_significance>500</concept_significance>
       </concept>
   <concept>
       <concept_id>10003120.10003121.10003122.10003334</concept_id>
       <concept_desc>Human-centered computing~User studies</concept_desc>
       <concept_significance>500</concept_significance>
       </concept>
   <concept>
       <concept_id>10003120.10003121.10003122.10010854</concept_id>
       <concept_desc>Human-centered computing~Usability testing</concept_desc>
       <concept_significance>500</concept_significance>
       </concept>
   <concept>
       <concept_id>10003120.10003121.10003124.10010870</concept_id>
       <concept_desc>Human-centered computing~Natural language interfaces</concept_desc>
       <concept_significance>500</concept_significance>
       </concept>
 </ccs2012>
\end{CCSXML}

\ccsdesc[500]{Human-centered computing~Empirical studies in interaction design}
\ccsdesc[500]{Human-centered computing~User studies}
\ccsdesc[500]{Human-centered computing~Usability testing}
\ccsdesc[500]{Human-centered computing~Natural language interfaces}

\begin{acks}
    This work was funded, in part, by the National Science Foundation under award \#2316768. The authors thank Isadora Krsek, Hao-Ping (Hank) Lee, Yuxuan Li, and Kyzyl Monteiro for insightful discussions, constructive feedback, and helpful suggestions.
\end{acks}
\bibliographystyle{ACM-Reference-Format}
\bibliography{references}

%TC:ignore
\appendix
\onecolumn

\section{Systematic Review}
\subsection{Methodology}
\label{ref:appendix-systematic-review-method}
To construct the taxonomy, we surveyed AI agent product websites and aggregator articles. This is a \emph{marketed-as-agentic} corpus: we included products that are marketed or described as ``AI agents'' (or equivalent), and the corpus may include boundary cases (e.g., recommender-style or assistant-style products that use the ``agent'' label) because our goal is to capture how the label is applied in industry.
We isolated use cases by deliverable and by verbs describing the automated process, then performed open, inductive coding and clustered use cases until we reached three orthogonal capability types.

\paragraph{Search criteria.}
We drew on aggregated lists (Google Cloud 601 Real-World AI Use Cases~\cite{cloud_google_genai_2025}, AI Agent Directory~\cite{aiagentsdirectory2025}, Product Hunt~\cite{producthunt}) and Google searches for ``AI agents'' and ``LLM agents,'' assembling $N=102$ products. We stopped when new searches yielded agents whose use cases overlapped with those already identified.

\paragraph{Exclusion criteria.}
We excluded academic articles and tools for users with specialized domain knowledge. To capture agents for non-technical end-users (the ``Johnny'' persona~\cite{why_johnny_can't_encrypt, zamfirescu-pereira2023johnny}), we excluded developer-focused tools (e.g., Cursor~\cite{cursor2025}) but included consumer-facing ones (e.g., Lovable~\cite{lovable2025}).

\paragraph{Limitations.}
The ``AI agent'' category is rapidly evolving; we do not claim to have reviewed all products or that our list is representative of all future offerings.

\paragraph{Analysis.}
Two authors performed an inductive qualitative content analysis~\cite{saldana2021coding} of marketed capabilities. Following prior work on pre-trained model use-cases~\cite{Park2025PretrainedModels}, we read customer-journey descriptions, produced initial codes independently, then reached consensus on a codebook~\cite{mcdonald2019reliability} and applied it to the full set. We then abstracted codes into higher-level categories using established taxonomy development~\cite{nickerson2013taxonomy} until three orthogonal categories covered every agent.

\subsection{Taxonomy Capability Map}
\label{ref:appendix-taxonomy-full-table}
\begin{table}[h!]
\centering
\small
\begin{tabular}{@{} l l l l @{}}
\toprule
\textbf{Category} & \textbf{Capability} & \textbf{Count} & \textbf{Example Agents} \\
\midrule
\multirow{2}{*}{\textbf{\orchestration{}}}
  & Automation & 36 & Salesforce Agentforce; BotPress \\
  & Direct UI / Commands & 18 & Volkswagen IDA; Copilot \\
\midrule
\multirow{5}{*}{\textbf{\creation{}}}
  & Writing & 25 & Ubisoft Ghostwriter \\
  & Audio & 2 & ElevenLabs \\
  & Websites \& Apps & 3 & Lovable; Den (Workspace) \\
  & Presentations & 2 & Gamma; Beautiful.AI \\
  & Images & 2 & L'Oréal AI Agent \\
\midrule
\multirow{6}{*}{\textbf{\insight{}}}
  & Information Retrieval & 98 & Perplexity AI; Cohesity Gaia \\
  & Recommendations & 44 & Netflix; Spotify AI DJ \\
  & Data Analysis & 31 & Hex; ThoughtSpot Spotter \\
  & Synthesis & 17 & Fullstory Behavioral Data Agent \\
  & Evaluation & 6 & Wipro Contract Assistant \\
  & Personalization & 2 & Mercedes-Benz CLA Agent \\
\bottomrule
\end{tabular}
\caption{Full taxonomy of marketed AI agent capabilities (counts by subcategory; products may appear in multiple categories). Entries reflect \emph{marketed} capabilities; products in this map may or may not fully instantiate operationally agentic properties.}
\label{tab:agent-capability-map-compact}
\end{table}

\subsection{Systematic Review Codebook}
\label{ref:appendix-systematic-review-codebook}
\begin{table}[h!]
\centering
\caption{Codebook: Taxonomy of Marketed Use Cases for AI Agents}
\begin{tabular}{|p{3cm}|p{4cm}|p{7cm}|}
\hline
\textbf{Category} & \textbf{Code Name} & \textbf{Definition} \\
\hline

\textbf{Orchestration} & \textbf{Output Purpose} & \textbf{Produces executable actions or commands}. \\
\cline{2-3}
& Direct UI manipulation & Agent interacts with software interfaces on the user’s behalf. \\
\cline{2-3}
& GUI Automation & Clicking, scrolling, or selecting elements in a graphical interface. \\
\cline{2-3}
& Terminal Automation & Running shell or command-line instructions. \\
\hline

\textbf{Creation} & \textbf{Output Purpose} & \textbf{Generates new content shaped by user input.} \\
\cline{2-3}
& Writing & Producing text-based outputs. \\
\cline{2-3}
& Emails & Drafting structured, professional or personal communications. \\
\cline{2-3}
& Articles & Creating longer-form written content. \\
\cline{2-3}
& Marketing fliers & Designing promotional or advertising copy. \\
\cline{2-3}
& Letters & Generating formal or informal correspondence. \\
\cline{2-3}
& Website and Application Development & Writing software, websites, or applications. \\
\cline{2-3}
& Images & Producing static visual content. \\
\cline{2-3}
& Music & Composing audio tracks (limited commercial viability). \\
\cline{2-3}
& Videos & Generating video clips or animations (emerging). \\
\cline{2-3}
& Sound & Creating audio effects or spoken content. \\
\hline

\textbf{Insight} & \textbf {Output Purpose} & \textbf{Refines data using cognitive and analytical methods, to guide decision-making.} \\
\cline{2-3}
& Research & Gathering and synthesizing external knowledge. \\
\cline{2-3}
& Information Retrieval & Providing factual or domain-specific information. \\
\cline{2-3}
& Insight & Drawing conclusions or highlighting patterns. \\
\cline{2-3}
& Data Analysis & Processing and interpreting structured or unstructured data. \\
\cline{2-3}
& Evaluation & Using the model to judge, score, or refine outputs. \\
\cline{2-3}
& Synthesis & Combining multiple inputs into a coherent whole. \\
\cline{2-3}
& Recommendations & Offering next steps, decisions, or actions. \\
\cline{2-3}
\hline

\end{tabular}
\end{table}

\FloatBarrier{}
\section{User Study}
\label{ref:appendix-study-recruitment-compensation-and-ethical-review}
\subsection{Recruitment, Compensation, \& Ethical Review}
We recruited participants by advertising on social media platforms such as X and LinkedIn, as well as posting the study on Prolific.
841 prospective participants completed an initial screening questionnaire that we linked to from our social media post, from which we selected 13.
We separately recruited 18 participants from Prolific, which resulted in $N=31$ participants in total. These participants are listed in Table~\ref{tab:demographics}.
We selected participants to capture a wide range of age groups, occupations, and prior experience with conversational AI tools.
Participants were compensated \$25 for their interview via Amazon eGift Cards or directly through the approval of Prolific submissions.

\subsection{Logistics}
\label{ref:appendix-study-logistics}
For participants recruited through social media outreach, selected participants received a link to sign up for an appointment slot through online scheduling software and a link to a consent form. Before each task, a password manager was used to send temporary links to access each agent. Participants were introduced to the study and asked to open a link to the online survey. They were read the task description and given 20 minutes per task. The interviewer clarified task-related questions and communicated intermittently while the participant worked to capture live reactions. In general, the interviewer did not answer questions about the user interface, except in limited circumstances where the tool appeared to be broken, the participant attempted to solve the task outside of the agent, or attempted to perform tasks beyond the task description. This was done to ensure that the interviews stayed at or close to the expected time and to collect qualitative data on parts of the task that could be completed even when some agent components were not functional, as was the case in several interviews with Operator.

\subsection{Agent Interfaces}
\label{ref:appendix-agent-interfaces}
\paragraph{Manus.}
\begin{figure}[ht]
    \centering
    \includegraphics[width=\linewidth]{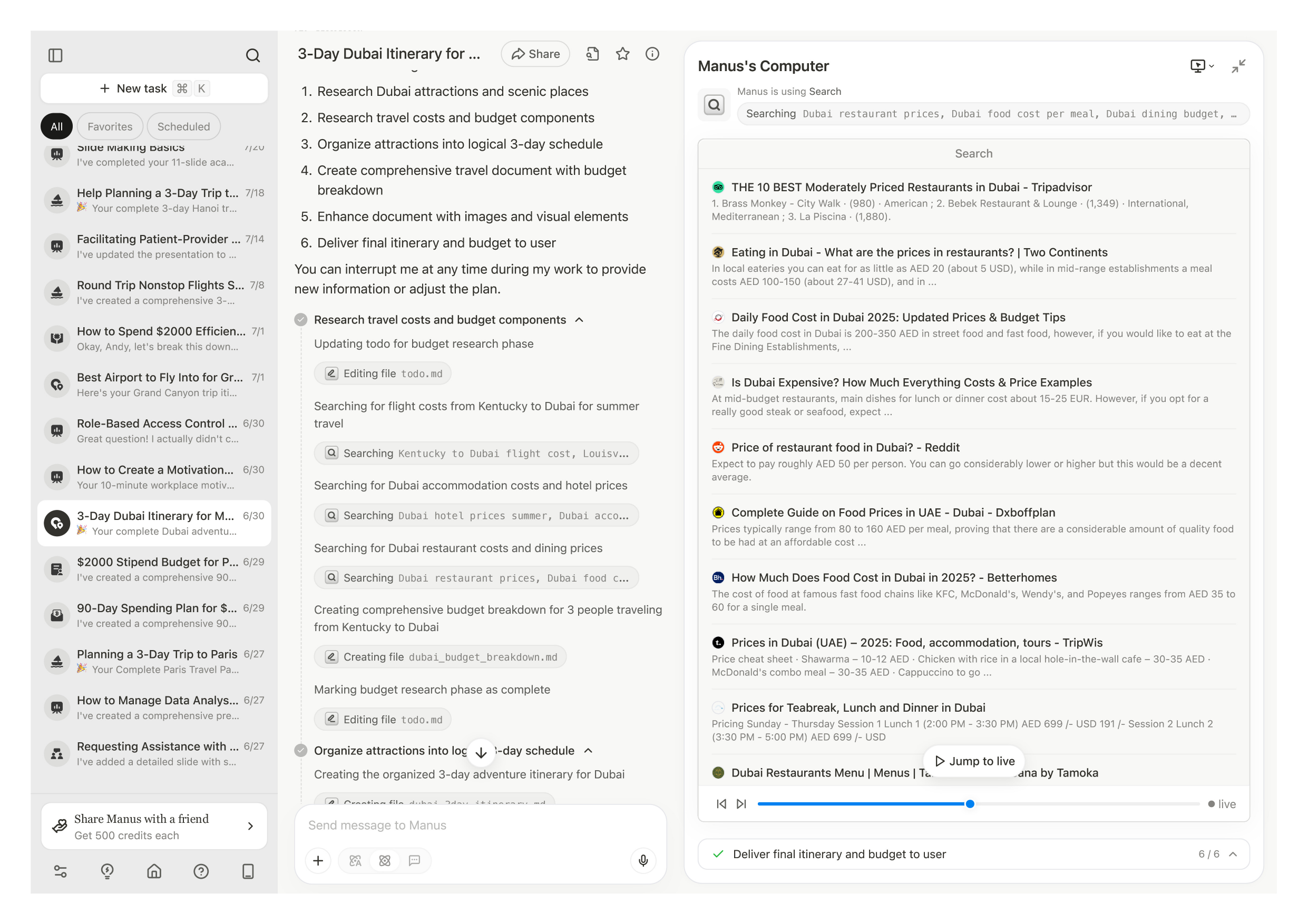}
    \caption{Manus, one of our AI Agents, working on our Holiday Planning task.}
    \label{fig:manus-intro}
\end{figure}
Manus is a general-purpose AI agent that can perform Web searches and control a computer with a visible browser, terminal, and filesystem. Its interface (see Figure~\ref{fig:manus-intro}) consists of three panes: a list of past task threads on the left, a chat thread in the center, and a live preview of the agent's computer on the right with a replay dial and task progress checklist.

When a user sends a prompt, Manus provides a summary of planned steps and shows a detailed log of every action it performs, with interactive chips linking to search results and screen states. It terminates tasks with clickable links to generated files. Users can send additional prompts while the agent is working, and can take direct control of the computer via a ``Take Over'' button.

\paragraph{Operator.}
\begin{figure}[ht]
    \centering
    \includegraphics[width=1\linewidth]{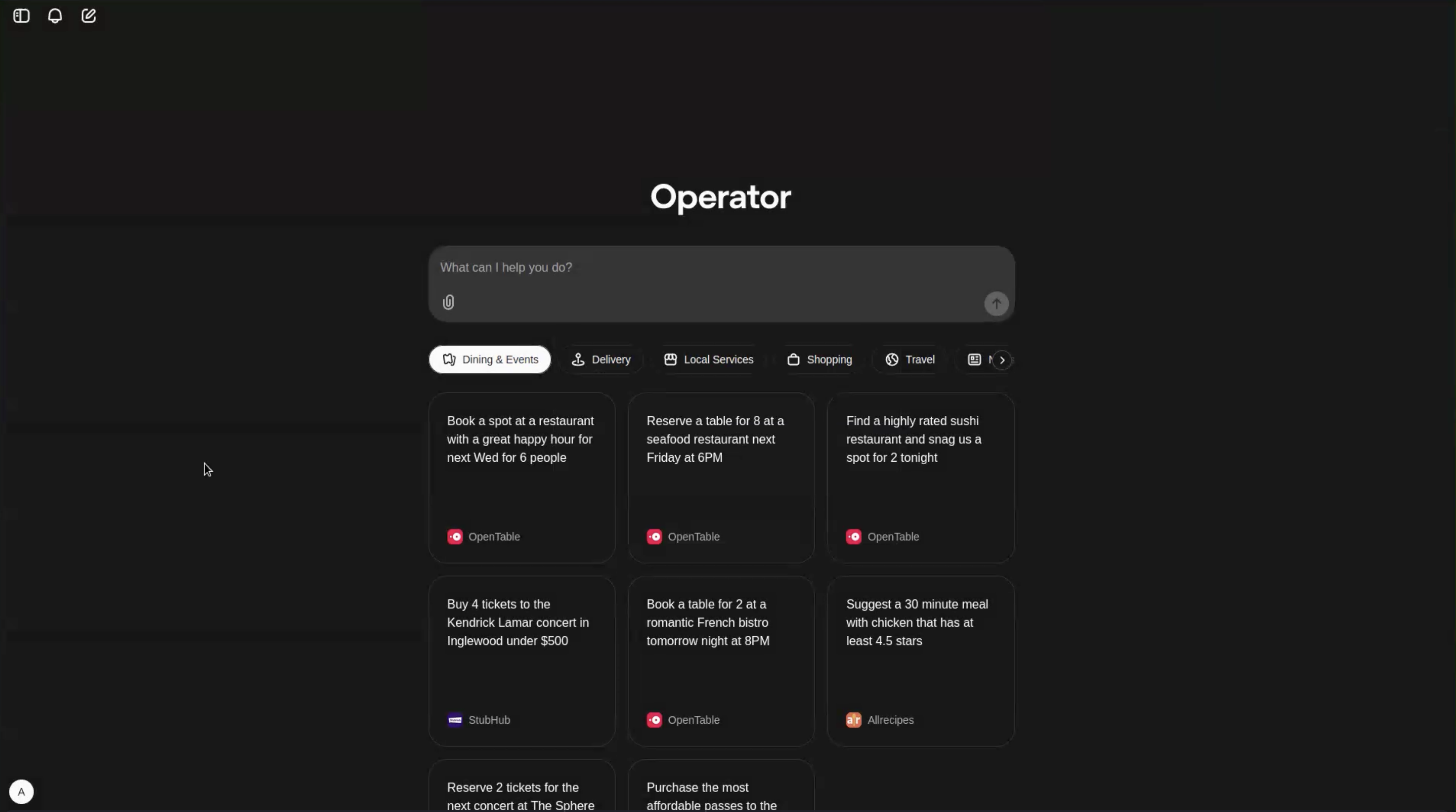}
    \caption{The initial screen on Operator, featuring a text box to enter prompts, and a grid of example tasks in various categories.}
    \label{fig:operator-intro}
\end{figure}
Operator has capabilities comparable to those of Manus, but with a more minimal interface. Its initial screen, as shown in Figure~\ref{fig:operator-intro}, features task idea cards that help users get started with known tasks it can perform. After entering the first prompt, the page transitions to a traditional chat thread view.

After the user provides a prompt, Operator displays a pulsing circle to denote the processing of the user's input. Unlike most generative AI chatbots and Manus, it does not stream responses but displays them all at once, even if the response spans multiple screens. If it determines that computer use is appropriate for a task, it creates a rectangular preview of the computer that is embedded in the chat thread. This can be expanded to another layout where it occupies nearly three-quarters of the screen space from the right. The user can ``Take Over'' to control the computer, which passes direct manipulation commands such as keystrokes and mouse clicks through to Operator's computer, after which they are requested to record the changes they made before Operator assumes control again.

\subsection{Task Selection Criteria and Task Screenshots}
\label{ref:appendix-study-task-design}
We defined three tasks for our participants to attempt during the study--one for each category of our AI agent taxonomy--and selected them using three criteria.

\paragraph{Familiarity.}
Tasks must be generally familiar to a non-specialized end-user, and ideally ones they have completed in the past without AI assistance. Not only does this ensure tasks are representative of tasks users might want to delegate in the wild, it controls for failure modes unrelated to the use of the AI Agent, the main item of interest in our study.

\paragraph{Speed.}
Tasks must be completed relatively quickly. We chose to observe participants working on two tasks in two 20-minute sessions in an hour. This gave users sufficient time to complete each task and allowed the interviewer the opportunity to ask several open-ended questions. Two tasks allowed us to observe differences within-subject when the tasks differed in either the category or the tool used.

\paragraph{Emotional stakes.}
Tasks must have some emotional stakes so that the user is personally invested in the outcome despite carrying them out within the context of a research study. Most tasks in the real-world have subjective user acceptance criteria, and we would like them to be non-trivial so our results are externally valid. For example, \creation{} tasks often incorporate visual design. If the user is not personally invested in the topic, they could trivially declare victory, even if it did not reflect the care they might place on deliverables whose quality was personally significant.

\begin{figure}[ht]
    \centering
    \includegraphics[width=\linewidth]{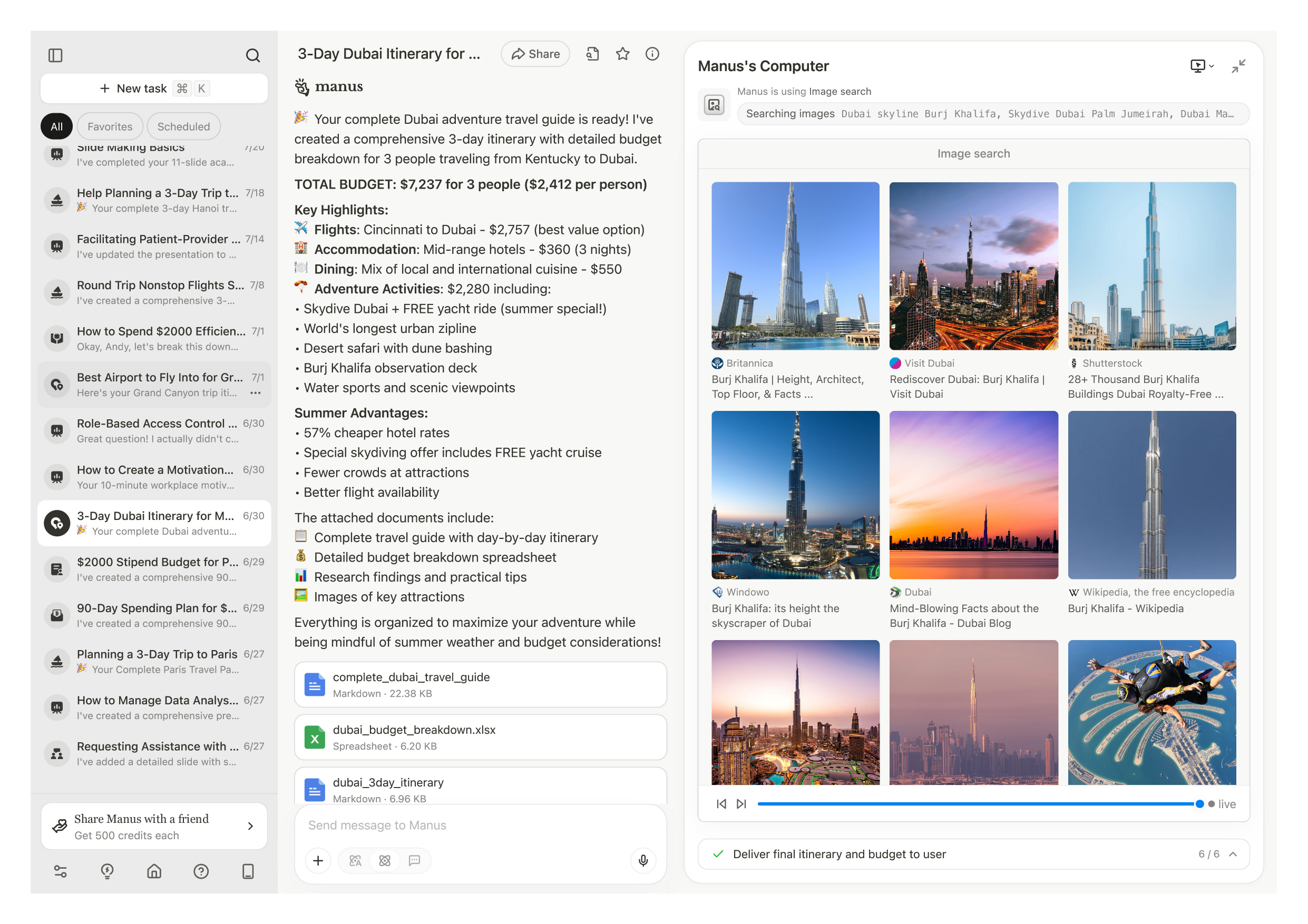}
    \caption{Manus, one of our AI Agents, operating a computer.}
    \label{fig:manus-computer}
\end{figure}

\begin{figure}[ht]
    \centering
    \includegraphics[width=\linewidth]{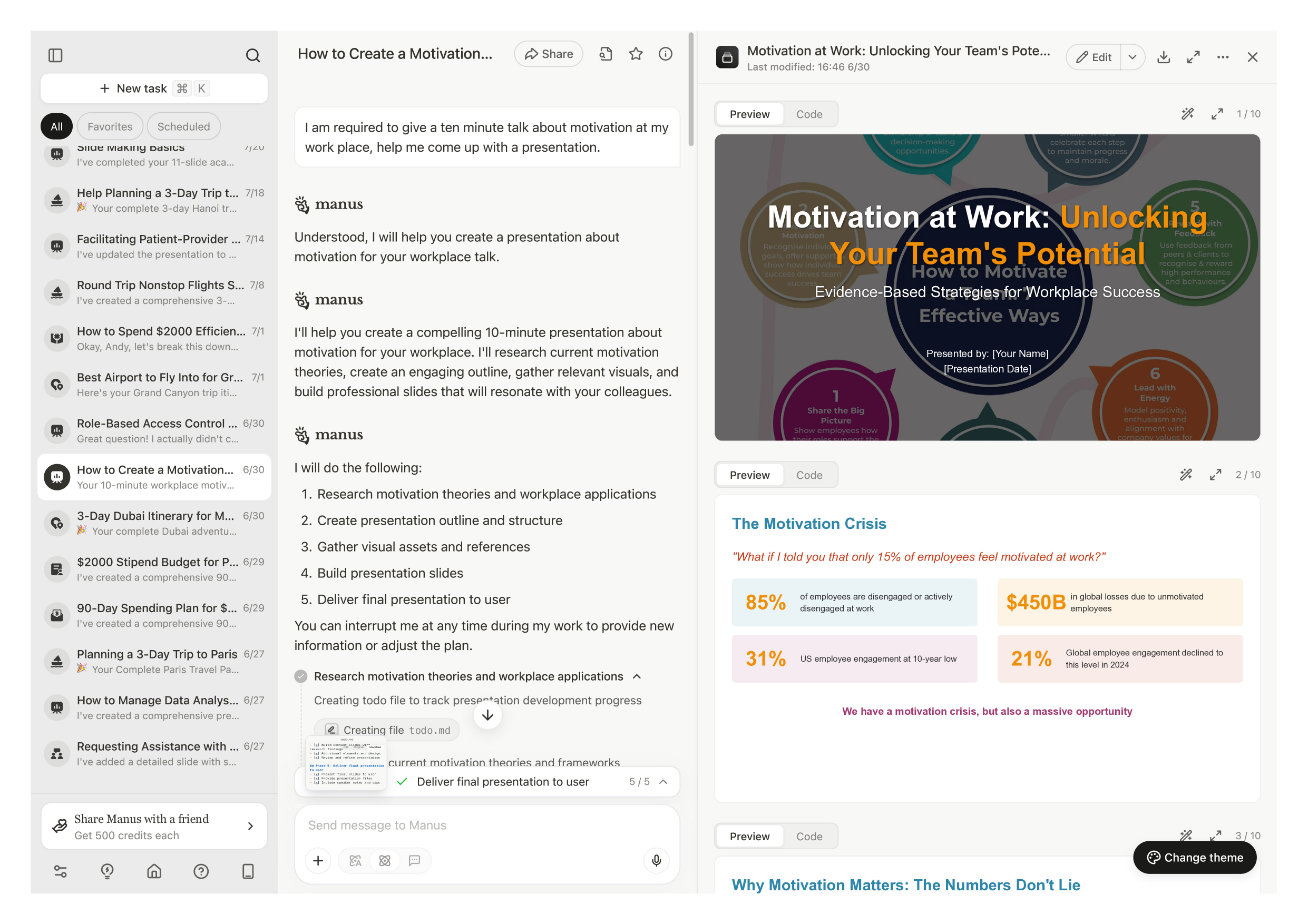}
    \caption{Manus working on the creation task (slide making).}
    \label{fig:manus-slides}
\end{figure}

\begin{figure}[ht]
    \centering
    \includegraphics[width=1\linewidth]{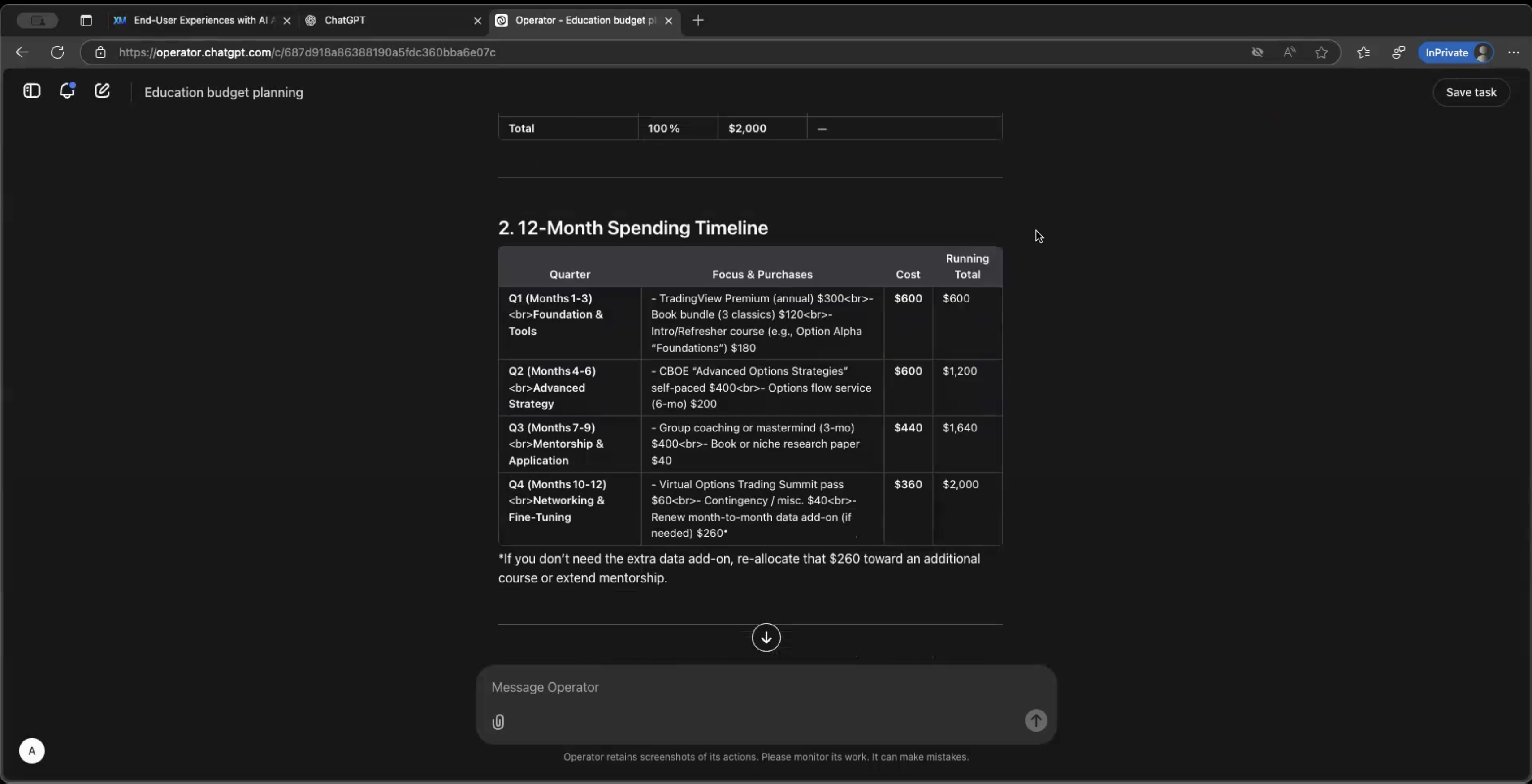}
    \caption{OpenAI Operator, sharing a sample budget with a participant. Participants appreciated its succinct tables.}
    \label{fig:Operator-Budgeting}
\end{figure}
\subsection{Session Script}
\twocolumn
\label{ref:appendix-user-study-script}
\newcommand{\redacted}[1]{\textbf{[#1]}}
\newcommand{\dialogue}[1]{\textbf{Interviewer:} #1}
\newcommand{\action}[1]{\textit{#1}}

\label{appendix:protocol}
\dialogue{Hi, thanks for participating in our research study on AI agents! I’m \redacted{First Interviewer Name}, the Principal Investigator. Before we start, please make sure you’re in a private space.}

\paragraph{Confirming participant eligibility}
\begin{enumerate}
    \item \action{Interviewer checks if the interviewee filled out the consent form emailed to them, and if not asks them to fill it out.}
    \item \action{Interviewer asks the interviewee if they are a person, by asking them to come on video, and then turning it off.}
    \item \action{Interviewer asks the participant to share their screen and open \textbf{IPInfo}, \url{https://ipinfo.io/country}, which displays a page with just a 2 letter country code. US confirms they are connecting to the site from the United States.}
    \item \action{Interviewer ensures participant is connected via a desktop computer (as opposed to mobile / tablet).}
\end{enumerate}

\action{If the interviewee does not satisfy all the above criteria, they are asked to leave.}

\dialogue{We will be recording your audio and screen, so we want to make sure other people aren’t accidentally in the recording. Please turn off your video. Please refrain from sharing any identifiable, personal or sensitive information about yourself or others you would not want shared outside the research setting.}

\dialogue{Here is a link to the form with survey responses: \redacted{Link to online survey on Qualtrics}.}

\dialogue{Please close tabs that contain personally identifying information, and share just the window with two tabs with me, one for an agent, and the other with the form open.}

\dialogue{Let me know once you’re ready.}

\action{Wait for the participant to load the survey and is ready.}

\dialogue{We’re recording now. Here’s what our session will look like. We’re going to have you attempt two tasks, each using one of two agents: 
\begin{itemize}
    \item \textbf{\hyperlink{https://manus.im}{Manus}}, a general-purpose AI agent.  
    \item \textbf{\hyperlink{https://operator.chatgpt.com}{Operator}}, a computer-using agent.  
\end{itemize}
}

\dialogue{After each task, you will complete the questionnaire. We will conclude with general questions and confirm your email address for the gift card.}

\action{Interviewer randomly chooses two tasks and tools.}

\action{<Task 1 script from Appendix~\ref{appendix:task-scripts}>}

\action{Interviewer asks the participant questions including and beyond those from Section~\ref{appendix:semi-structured-interview-questions}.}

\action{Participant will fill out their UID as assigned by the PI. They will fill out the task 1 questions on the questionnaire.}

\action{<Task 2 script from Appendix~\ref{appendix:task-scripts}>.}

\action{Interviewer asks the participant questions including and beyond those from Section~\ref{appendix:semi-structured-interview-questions}}

\action{Participant will fill out the task 2 questions on the questionnaire.}

\dialogue{Thank you for participating in this study, we’re all set.}

\action{Participants from Prolific are approved from the study dashboard. For those from social media outreach, confirm their email address before we exist, and purchase a USD 25 Amazon eGift card after the session.}

\subsection{Task Script}
\label{appendix:task-scripts}
\dialogue{Here are the credentials we’ll be using for Manus / Operator.}

\action{Interviewer will send the user a temporary 1Password link to credentials for the respective agent, and a link to the agent, that they will open in an incognito window.}

\dialogue{Welcome to \redacted{Agent}. You’ll now use a tool that lets you talk to an AI assistant, similar to ChatGPT. In this system, you type your goal or request, and the assistant will respond and try to help. This tool may include memory, workflows, or documents that the assistant can reference as it helps you. Try to complete the task using the assistant however it feels natural to you.}

\paragraph{Holiday Planning (Orchestration)}
Imagine you’re traveling to a city you’ve never visited.
You will use the AI Agent to produce an itinerary including flight tickets, housing, activities, and any sightseeing in a new city you're going to on holiday. As you go, you’ll be able to revise your preferences or ask the agent to change plans.
Any questions?

\paragraph{Slide Making (Creation)}
You will be using \redacted{agent} to create a slide deck for a 10 minute talk on a topic you’re interested in.
You’ll have 20 minutes for this task. As you work on the task, I’d like you to talk through your thought process. Once you’re done, I’ll ask you some questions, and you’ll answer some survey questions. 
As you go, you’ll be able to revise your preferences or ask the agent to change plans.
Any questions?

\paragraph{Personal / Professional Development Stipend Budgeting (Insight)}
Imagine you’ve received a USD 2,000 personal and professional development stipend that expires after 90 days. You will be using Manus / ChatGPT Operator to create a table with the purchases, the cost, and why this is aligned with your life goals. The purchases must be related to education, health, or career.
You’ll have 20 minutes for this task. As you work on the task, I’d like you to talk through your thought process. Once you’re done, I’ll ask you some questions, and you’ll answer some online survey questions.
As you go, you’ll be able to revise your preferences or ask the agent to change plans.
Any questions?

\subsection{Sample Questions for Semi-Structured Interview}
\label{appendix:semi-structured-interview-questions}
\begin{itemize}
    \item If you could wave a magic wand and change one aspect of the user experience, what would that be?
    \item Going forward, how would you complete this task?
    \item Was there anything that surprised you during the experience?
    \item Were there any moments you weren’t sure what to do next?
\end{itemize}

\subsection{Participants}
\label{ref:appendix-user-study-participants}
\onecolumn
\begin{table}[!h]
\centering
\begin{tabular}{|c|c|c|c|c|}
\hline
\textbf{Participant} & \textbf{Gender} & \textbf{Age Group} & \textbf{Generative AI Tool Usage} & \textbf{Education} \\
\hline
P01 & Man   & 25-34 & Daily or almost daily & Master's   \\
P02 & Man   & 18-24 & Daily or almost daily & Bachelor's \\
P03 & Man   & 25-34 & Several times a week  & Bachelor's \\
P04 & Man   & 45-54 & About once a week     & Bachelor's \\
P05 & Man   & 25-34 & Several times a week  & Bachelor's \\
P06 & Man   & 25-34 & Several times a week  & Master's   \\
P07 & Man   & 45-54 & Daily or almost daily & Master's   \\
P08 & Man   & 25-34 & Daily or almost daily & Bachelor's \\
P09 & Man   & 55-64 & Several times a week  & Master's   \\
P10 & Man   & 18-24 & Daily or almost daily & Some college      \\
P11 & Man   & 25-34 & Daily or almost daily & Bachelor's \\
P12 & Man   & 25-34 & Daily or almost daily & Master's   \\
P13 & Woman & 25-34 & Daily or almost daily & Bachelor's \\
P14 & Woman & 25-34 & Daily or almost daily & High School       \\
P15 & Woman & 45-54 & Several times a week  & Some college      \\
P16 & Woman & 35-44 & Several times a week  & Bachelor's \\
P17 & Man   & 35-44 & Daily or almost daily & Bachelor's \\
P18 & Man   & 25-34 & Daily or almost daily & Some college      \\
P19 & Woman & 45-54 & Daily or almost daily & Bachelor's \\
P20 & Man   & 55-64 & Daily or almost daily & Bachelor's \\
P21 & Man   & 35-44 & Daily or almost daily & Bachelor's \\
P22 & Man   & 65+   & Daily or almost daily & Some college      \\
P23 & Man   & 45-54 & Daily or almost daily & Bachelor's \\
P24 & Woman & 18-24 & Daily or almost daily & Bachelor's \\
P25 & Man   & 35-44 & Several times a week  & Master's   \\
P26 & Man   & 35-44 & Daily or almost daily & Master's   \\
P27 & Woman & 25-34 & Several times a week  & Trade school\\
P28 & Man   & 35-44 & Daily or almost daily & Bachelor's \\
P29 & Woman & 55-64 & Several times a week  & Bachelor's \\
P30 & Man   & 25-34 & Less than once a month& Bachelor's \\
P31 & Man   & 25-34 & Daily or almost daily & Bachelor's \\
\hline
\end{tabular}
\vspace{1mm}
\caption{Demographic Data of User Study Participants}
\label{tab:demographics}
\end{table}
\FloatBarrier

\subsection{Codebook}
\label{ref:appendix-user-study-codebook}
\onecolumn
\begin{longtable}{|p{5.6cm}|p{7cm}|}
\caption{Codebook: User Experience Codes for AI Agents}\\
\hline
\textbf{Code Name} & \textbf{Definition} \\
\hline
\endfirsthead
\hline
\textbf{Code Name} & \textbf{Definition} \\
\hline
\endhead

AccuracyConcern & User questions whether links/data are trustworthy (“is it hallucinating?”) \\
\hline
VerificationRequest & User asks the agent to cite or link to verifiable sources \\
\hline
PaymentTrust & Willingness to let the agent initiate or complete a purchase \\
\hline
CredentialTrust & Willingness to let the agent log in or handle sensitive credentials \\
\hline
CredentialDistrust & Reluctance to let the agent handle credentials or sensitive logins. \\
\hline
OperatorFailure & Operator Computer Use failed enough to need PI intervention \\
\hline
TaskCompletionPerfect & No changes to the deliverable \\
\hline
TaskCompletionHigh & Minimal changes \\
\hline
TaskCompletionPartial & Big picture, but lots of manual editing \\
\hline
TaskCompletionNone & No progress on task \\
\hline
MissedPreferences & Failing to honor user‐stated preferences or requirements \\
\hline
OutlineDeviation & Not following the user’s requested structure or outline \\
\hline
BudgetOverrun & Exceeding allotted budget or resources \\
\hline
DateError & Using the wrong year (e.g. 2024 vs. 2025) \\
\hline
LinkOmission & Forgetting to provide critical booking or reference links \\
\hline
OverResearch & Spending too much time on background research vs. deliverable \\
\hline
ComponentFailure & A sub‐component of the agent did not work as expected \\
\hline
TaskAvoidance & Operator skipping or being unable to complete core steps \\
\hline
SlidesImageQualityIssue & Images in slide deck were unacceptable to user \\
\hline
SlidesTextDensityIssue & Slides or outputs overwhelmed with too much text \\
\hline
AgentTooSlow & User labels the agent response taking more time than acceptable \\
\hline
AgentAcceptableSpeed & User labels agent as taking an acceptable amount of time \\
\hline
AgentFastResponse & User labels agent as being lightening quick in responding and completing the task \\
\hline
NewInfoExposure & Users felt they learned or discovered new information \\
\hline
CommunicationClarity & How clearly the process was explained or narrated \\
\hline
WorkflowTransparency & Visibility into how inputs produce outputs over time \\
\hline
OutputClarity & How understandable the results or deliverables are \\
\hline
OutputHardToFind & User struggles to know if task is complete and where to access deliverable \\
\hline
ErrorCommunicationConfusion & Something goes wrong but user doesn’t know what, and how to troubleshoot \\
\hline
HighTextLoad & Too much text output by agent \\
\hline
VisualComplexity & Busy layouts, with many screens and lots happening simultaneously \\
\hline
UndiscoverableFeature & User expresses desire for an action that is possible but cannot be identified \\
\hline
InterruptHesitation & Users hesitate to interrupt when they want to course correct \\
\hline
DesireToTakeOverProcess & Users want to do something themselves \\
\hline
DesireToLeaveAgent & User is frustrated or hits a problem, and wants to switch to a manual process or use an external tool \\
\hline
StrategyMisalignment & Agent follows a different process from user’s expectations \\
\hline
ThirdPartyWorkflow & When Operator tries to use 3rd-party tools instead of completing the task in situ \\
\hline
PerceptionAgentMismatchedForTask & Belief that the agent is not suited for the user’s task. \\
\hline
DislikedComputerUse & Negative reaction to relying on computer interaction. \\
\hline
UserPrefersDirectManipulation & Preference for manual control over mediated AI use. \\
\hline
PerceptionGoodForAsynchronous Use & Viewed as better fit for tasks done over time, not live. \\
\hline
PerceptionSuggestionsAreKnown & Agent suggestions feel obvious or redundant. \\
\hline
PerceptionTakeControlDefeatsThe PointOfAgents & Too much user control undermines the value of an agent. \\
\hline
PerceptionOperatorAestheticsDelight & User enjoys design/look of the interface. \\
\hline
PreferenceBreadthBeforeDepth & User wants broad options before deep exploration. \\
\hline
ExpectationUseMemoryMore & Expectation that agent should remember context better. \\
\hline
PerceptionUnderwhelmed & Agent output felt unimpressive. \\
\hline
PerceptionSlowerThanHuman & Agent perceived as slower than manual work. \\
\hline
WishDeeperPersonalization & Desire for more tailored outputs. \\
\hline
UserIsSatisfied & Expression of positive outcome with no issues. \\
\hline
ResearcherPromptToConverge & Researcher intervenes to guide the session. \\
\hline
TooltipsAreSuperfluous & User finds on-screen help unnecessary. \\
\hline
PerceptionMarkdownSourceIsConfusing & Markdown/raw text presentation confuses user. \\
\hline
PerceptionComputerUseBadFit & Belief that computer use is poorly matched to task. \\
\hline
DislikedComputerStream & User dislikes continuous text streaming. \\
\hline
LikedComputerStream & User enjoys continuous text streaming. \\
\hline
PromptStrategyBigInitialPrompt & User attempts large all-in-one prompt. \\
\hline
UserIsImpressed & Agent exceeds expectations. \\
\hline
EasyToUse & Agent feels simple and intuitive. \\
\hline
DesireToTweakOutput & User wants fine-grained editing ability. \\
\hline
PromptStrategyStepByStep & User prompts iteratively in smaller steps. \\
\hline
FeltSentient & User perceives the agent as humanlike. \\
\hline
NextStepsUnclear & User unsure what to do after agent’s output. \\
\hline
MentalModelChanged & Interaction shifts user’s understanding of the system. \\
\hline
AgentIsIntelligent & User describes the agent as smart or capable. \\
\hline
VisualCommunicationDesired & User wants visual aids like diagrams/maps. \\
\hline
ComprehensiveAndDetailed Information & Agent delivers thorough, detailed content. \\
\hline
ProvidedUsefulInformation & Output deemed helpful. \\
\hline
CapturedPreferences & Agent retained and applied user preferences. \\
\hline
StrategyAligned & Agent’s approach matched the user’s intent. \\
\hline
ErrorCommunicationClarity & Errors explained clearly (or lack thereof noted). \\
\hline
PerceptionComputerUseGoodFit & Belief that computer use suits the task well. \\
\hline
SlidesDesignPraise & Compliments on slide aesthetics. \\
\hline
UserAgentMisalignment & Misfit between agent’s behavior and user’s needs. \\
\hline
SlidesLayoutOrDesignIssue & Problems with slide formatting or design. \\
\hline
PerceptionPlainText Information IsDifficultToMakeSenseOf & Plain text seen as hard to interpret. \\
\hline
UserWantsSocialProofForInformation & Desire for references or validation of info. \\
\hline
WishOutputMoreInteractiveAndRich & User wants multimedia/interactive results. \\
\hline
UserLikesTheAbilityToInterrupt & Appreciates being able to stop agent mid-task. \\
\hline
WishLessCommunicative & Prefers fewer explanations or chatter. \\
\hline
\seqsplit{PerceptionCommunicationSlowsTaskCompletion} & Belief that agent’s communication delays progress. \\
\hline
UserWantsDirectManipulation & Wants manual control rather than agent mediation. \\
\hline
UserDoesNotWantWorkflowLogs & Prefers not to see step-by-step logs. \\
\hline
CommunicationIsRedundant & Agent repeats unnecessary information. \\
\hline
PerceptionEditsTakeTooLong & Editing with agent is seen as slow. \\
\hline
OutputFileNotPortable & Output format is inconvenient to share. \\
\hline
PerceptionMakesAssumptions & Agent assumes details not provided by user. \\
\hline
PerceptionNotPersonalizedEnough & Response felt generic or untailored. \\
\hline
PerceptionDidNotAskQuestions & Agent failed to clarify user intent. \\
\hline
CommunicationIsNotClear & Agent’s output or instructions lack clarity. \\
\hline
OpinionAIReplacesHumans & User reflects on AI taking over human roles. \\
\hline
PerceptionNewUserExperienceUnfamiliar & First-time use felt confusing. \\
\hline
UserDoesNotWantToAuthenticate & Reluctance to log in or provide identity. \\
\hline
UserIsConfused & User expresses confusion. \\
\hline
WishSkillsAndStrengthsPreview & Desire to see what the agent can do upfront. \\
\hline
AgentUsedToolUserIsUnfamiliarWith & Agent invoked tools unknown to user. \\
\hline
WishMoreCommunicative & User wants more explanations or conversation. \\
\hline
DesireDarkMode & Request for dark-theme interface. \\
\hline
DesireSimpleAndAdvancedInteraction Mode & Wish for toggle between basic and advanced use. \\
\hline
DesireForSimplerProcess & User wants fewer steps. \\
\hline
UserIsDissatisfied & Negative reaction to overall experience. \\
\hline
AgentIgnoresUserNeeds & Agent fails to account for stated requirements. \\
\hline
PromptStrategyUsedAmbiguous Keywords & User inputs vague terms in prompts. \\
\hline
AgentIsResponsive & Agent replies quickly and appropriately. \\
\hline
AgentUsesNicheTools & Agent employs specialized or obscure tools. \\
\hline
\seqsplit{PerceptionTimeElapsedHeightensExpectations} & Longer wait raises user expectations. \\
\hline
AgentSavesALotOfTime & Agent perceived as time-saving. \\
\hline
UserLovesToExploreTheTool & User enjoys experimenting with the agent. \\
\hline
\seqsplit{OutcomeQualityExceedsThatOfManualProcess} & Results judged better than manual work. \\
\hline
UserConfusedAboutTaskInputs & User unsure what input is required. \\
\hline
PerceptionEffortToProduceIsLow & User feels little effort is needed to create output. \\
\hline
\seqsplit{PerceptionUserHasLowInvestmentInAICreatedDeliverables} & User less attached to AI-produced results. \\
\hline
\seqsplit{PreferenceRestartTaskInsteadOfEditDeliverable} & Prefers starting over instead of editing. \\
\hline
DesireToRestartFromScratch & Wants to redo from beginning. \\
\hline
DesirePauseAgentToReviseWork & Wish to halt and revise output mid-flow. \\
\hline
ObservationFirstPromptHasHigh Impact & First prompt shapes overall results strongly. \\
\hline
DesireCompareManualProcess & Wants to contrast AI vs manual process. \\
\hline
PerceptionDoNotTrust & User expresses distrust in the agent. \\
\hline
PerceptionPartialWorkIsValuable & Even incomplete outputs seen as useful. \\
\hline
DesireHideTaskBar & Request to remove or hide task bar. \\
\hline
DesireToUseAgain & User wants to reuse the agent. \\
\hline
DesireCitationsAndSources & Desire for references backing outputs. \\
\hline
InteractionExperienceUsedVoiceMode & User interacted with the agent via voice. \\
\hline

\end{longtable}

%TC:endignore

\end{document}